%% file: Fransquinietalrevised.tex
\documentclass[preprint,12pt,authoryear]{config/elsarticle}

\input{packages}
\usepackage{multirow}
\begin{document}

\newcommand{\mariah}[1]{{\color{black}#1}}
\newcommand{\rodrigo}[1]{{\color{black}#1}}

\begin{frontmatter}

\title{Community-based anomaly detection using spectral graph filtering}


\author[1]{Rodrigo Francisquini}
\cortext[cor1]{Corresponding author}
\ead{rodrigo@francisquini.com}

\author[2]{Ana Carolina Lorena}
\ead{aclorena@ita.br}

\author[1,2]{Mariá C. V. Nascimento\corref{cor1}}
\ead{mcv.nascimento@unifesp.br;mariah@ita.br}

\address[1]{Instituto de Ciência e Tecnologia, Universidade Federal de São Paulo (UNIFESP), Av. Cesare M. G. Lattes, 1201, Eugênio de Mello, São José dos Campos-SP,
CEP: 12247-014, Brazil}

\address[2]{Divisão de Ciência da Computação (IEC), Instituto Tecnológico de Aeronáutica (ITA), Praça Marechal Eduardo Gomes, 50, Vila das Acácias, São José dos Campos-SP, CEP 12228-900, Brazil}

\begin{abstract}
Several applications have a community structure where the nodes of the same community share similar attributes. Anomaly or outlier detection in networks is a relevant and widely studied research topic with applications in various domains. Despite a significant amount of anomaly detection frameworks, there is a dearth on the literature of methods that consider both attributed graphs and the community structure of the networks. \mariah{This paper proposes a community-based anomaly detection algorithm using a spectral graph-based filter that includes the network community structure into the Laplacian matrix adopted as the basis for the Fourier transform. In addition, the choice of the cutoff frequency of the filter considers the number of communities found.  In computational experiments, the proposed strategy, called \textit{SpecF}, showed an outstanding performance in successfully identifying even discrete anomalies. }\textit{SpecF} is better than a baseline disregarding the community structure, especially for networks with a higher community overlapping. Additionally, we present a case study to validate the proposed method to study the dissemination of COVID-19 in the different districts of São José dos Campos, Brazil.
\end{abstract}




\end{frontmatter}

\section{Introduction}

The explosive growth of technology has led to a substantial increase in the amount of data collected from several applications. They include sensor measurements, transportation, internet, biological data, financial transactions, among others. Therefore, data analysis and processing techniques to manage massive amounts of data are of paramount importance. Data structure is usually irregular  and  relational data is commonly represented through graphs or networks \citep{ma2021comprehensive, dong2019learning}. 

\mariah{A great number of applications consists of networks with community structure. Some examples are in internet of things (IoT) \citep{chen2021learning}, protein-protein interactions \citep{francisquini2021community} and COVID-19 related data \citep{francisquini2021community}. Anomaly detection in this type of network is a specially relevant task. Anomaly detection in networks with community structure consists of identifying nodes, called anomalous nodes, that significantly differ from the standard observed in their community they belong to \citep{Akoglu2015GraphSurvey}. These are also known as context anomaly.

Despite being the target of intense investigation, there are networks for which graph anomaly detection tools are limited. There is a dearth of literature on data anomaly approaches that deal with data with relational properties and temporal information. Moreover, the topological graph structure is not usually taken into consideration on time series graph anomaly detection algorithms. According to Chen et al. \cite{chen2021learning}, applications such as sensor networks, for which there is a strong geographical and  temporal dependency, can benefit from frameworks that consider 
the topological graph structure.

Anomaly detection in networks has  attracted a lot of attention in the last five years with a soar on the number of studies  \citep{ma2021comprehensive}.  The reason behind this phenomena is not only the increase on the amount of applications but also the advent of sophisticate deep learning tools to perform such a task \citep{ma2021comprehensive, chen2021learning}. 
Most of the existing deep learning-based methods to time series anomaly detection uses semi-supervised learning, requiring some labeled data. In addition, according to Choi et al. \cite{choi2021deep}, the existing methods are too case-specific, demanding domain knowledge. 

The goal of this paper is to give new insights into anomaly detection in networks by designing a more generic unsupervised anomaly detection tool to approach attributed networks with community structure. These networks model a wide range of applications as the attributes of the networks are not limited to time series. As a result, for example, networks representing biological processes can be considered by the proposed method \cite{francisquini2021community}.  Moreover, this paper introduces a framework that considers the topological graph structure to describe the anomalies.  To this end, the proposed algorithm is founded on a recent signal processing concept which has also been drawing the attention of the data analysis research community, the \textit{graph signal processing} \citep{ortega2018graph}. }

Graph signal processing (GSP) techniques extend concepts from classical signal processing to signals indexed by generic graphs \citep{AliakseiSandryhaila2013, Shuman2013} and seek to analyze the data considering its underlying relational structure. Data from various domains, such as sensor networks, molecular network interactions, financial transactions, can be modeled as graph-indexed signals. For example, graphs can represent data collected from sensor networks, where sensors correspond to vertices, and edges connect sensors close to each other. The signals on the graph nodes correspond to the set of values measured by the sensors at a given time. Thus, GSP tools are used for many purposes, such as fault diagnosis, signal denoising, signal compressing, and anomaly detection \citep{Chen2014, Shuman2013,gao2021fault}.

Two approaches are commonly employed for processing signals indexed by graphs. The first considers the Laplacian matrix and is based on the spectral graph theory \citep{ChungFanRKandGraham1997}. The second relies on the adjacency matrix and is based on algebraic signal processing theory \citep{Puschel2008, Puschel2008a}. Both approaches generalize classical signal processing operations, such as filtering and Fourier transform, to the graph domain, by defining the concept of graph filters and graph Fourier transform, respectively \citep{Sandryhaila2014}. 

This paper proposes a method to detect anomalies in signals indexed by graphs using GSP theory and spectral graph theory.  In comparison to the related literature, the introduced method, named \textit{SpecF}, not only considers the adjacency relationships between vertices but also takes into account the community structure in the graph Fourier transform. This is achieved by incorporating the network community structure into an expanded adjacency matrix. The expanded adjacency matrix can be understood as a modification of the original unweighted graph to a weighted network through the inclusion of new edges. The method marks vertices whose signal values are outside the expected behavior for the community they belong as potential anomalies.

Computational experiments comparing the accuracy of the novel method with a counterpart using the classical adjacency matrix evidence the superiority of our community detection-based strategy 
in recovering anomalies, even for the most discrete cases. \mariah{In addition to experiments with labeled artificial networks proposed in this paper,  a comparative analysis with state-of-the-art time series anomaly detection algorithms on two IoT databases publicly available is performed. }This paper also shows an experiment with the proposed strategy on a COVID-19 dataset. The results of this experiment indicates an anomalous growth in the number of COVID-19 cases in the different districts of an upstate city of São Paulo, Brazil. 
\mariah{The main contributions of this paper are presented next.

\begin{itemize}
  \item We propose an 
  anomaly detection algorithm that addresses attributed networks for which the literature is scarce of anomaly detection algorithms;
    \item We introduce an 
    anomaly detection algorithm,  \textit{SpecF}, 
    based on graph signal processing theory which defines anomalous objects as those nodes whose signals required the most significant correction by a low-pass filter;
     \item We propose an anomaly detection algorithm, \textit{SpecF}, which differs from  other community-based anomaly detection methods by explicitly using the community structure in an extended adjacency matrix; 
    \item We introduce a 
    methodology to add normal and anomalous signals to networks with community structure, to better assess the proposed framework;
    \item We apply \textit{SpecF} to COVID-19 data to analyze the dissemination of the COVID-19 virus by investigating the anomalous districts of an upstate city from Brazil with approximately 730 thousand inhabitants.
\end{itemize}
}

\mariah{The remainder of this paper is organized as follows. Section~\ref{sec:gp} presents fundamental concepts of the GSP theory relevant for the understanding of the proposed tool. Section~\ref{sec:rw} shows a brief literature review on related anomaly detection algorithms. Section~\ref{sec:ad} introduces the proposed anomaly detection algorithm, \textit{SpecF}. Section~\ref{sec:ce} presents the computational experiments, including a thorough analysis of the anomaly detection algorithm in the COVID-19 dataset compiled in this study. Finally,  Section~\ref{sec:c} presents  final remarks and future research directions.}

\section{Graph Signal Processing}\label{sec:gp}
Data from several applications can be represented by graphs. Let $G = (V,E, A)$ be a weighted graph, where $V$ and $E$ are its respective sets of $n$ nodes and $m$ edges, and $A\in \mathbb{R}^{n\times n}$ is the weighted adjacency matrix. Each node $v_i\in V$ describes an instance of the dataset, and the weight $a_{ij}$ of an undirected\footnote{This paper assumes that  $G$ is an undirected graph.} edge carries the strength of the relation between a pair of vertices $v_i$ and $v_j$. The degree of a vertex $v_i$ is quantified by the sum of the weights of the edges incident to it. Let $D\in\mathbb{R}^{n\times n}$ be a diagonal matrix, called  degree matrix, where its $i^{th}$ diagonal
element $d_{ii}$ receives the degree of node $v_i$. Moreover, we denote here the set of neighbors of a vertex $v_i$ by $\mathcal{N}_i$, which means that this set contains all vertices adjacent to $v_i$.


The graph signal, defined by function $f:V \rightarrow \mathbb{R}$, is represented by a vector $\mathbf{f} \in \mathbb{R}^n$, where each element $f_i$ corresponds to the signal of node $v_i\in V$, i.e, $f(v_i)$.

\subsection{Graph Fourier Transform}\label{sec:fourier}

The graph Laplacian, also known as the non-normalized graph Laplacian, is the matrix $L = D - A$, corresponding to a real-valued symmetric matrix \citep{VonLuxburg2007AClustering}. Let $U=[u_{ij}]_{n\times n}$  be the set of eigenvectors of $L$.  \mariah{Without loss of generality, let $G$ be a connected component. Therefore $U$ is orthonormal, since $L$ is a real-valued symmmetric matrix. Moreover, the associated non-negative eigenvalues are referred here to as ${\{ \lambda_l \}}_{l=0, 1, ..., n-1}$, where $ 0 = \lambda_0 < \lambda_1 \leq \lambda_2 ... \leq \lambda_{n-1} \coloneqq \lambda_{max}$. As $U$ is orthornomal,  $U^TU=I$ holds.}


 The graph Fourier transform $\hat{\mathbf{f}}$ of a signal $\mathbf{f} \in \mathbb{R}^n$ given a symmetric shift operator $S=R \Lambda R^*$ is $\hat{\mathbf{f}}=R^*\mathbf{f}$ \citep{Shuman2013}. As $L=U \Lambda U^T$, $U^*=U^T$, the graph Fourier transform given $L$ is calculated by $\mathbf{\hat{f}}=U^T\mathbf{f}$. In addition, the $l$-th row of $U^T$ corresponds to the eigenvector associated with $\lambda_l$ and each component of $\mathbf{\hat{f}}$ can be written as:

\begin{equation}\label{eq:fourier}
     \hat{f}(\lambda_l)  = \sum_{i=1}^{n} u_{il} f(v_i), \ \ \ \forall l\in \{0,\ldots,n-1\}
 \end{equation}

Let $\mathbf{i\hat{f}}=U\mathbf{\hat{f}}$ be the inverse graph Fourier transform of $\mathbf{\hat{f}}$ considering the same shift operator $L$. It is possible to return to the signal $\mathbf{f}$ by the inverse graph Fourier transform of $\mathbf{f}$, since $$\mathbf{i\hat{f}}=U\mathbf{\hat{f}}=U U^T\mathbf{f}=I\mathbf{f} = \mathbf{f}$$ Therefore, we have that

\begin{equation}\label{eq:inversefourier}
     f(v_i)  = \sum_{l=0}^{n-1} u_{il} \hat{f}(\lambda_l), \ \ \ \forall i\in \{1,\ldots,n\}. 
 \end{equation}

\subsection{Frequencies on Graphs}

In classical Fourier analysis, eigenvalues \textbf{$\{(2\pi\xi)^2\}$} carry the notion of high and low frequencies. Low frequencies are associated with smooth complex exponential eigenfunctions that oscillate slowly, whereas high frequencies are related to complex exponential eigenfunctions that oscillate more rapidly.

\mariah{The definition of high and low frequencies for signals indexed by graphs takes into account graph Fourier theory (GFT). According to the GFT, the eigevectors of graph Laplacians associated to the first eigenvalues vary more sharply. Consequently, the eigenvectors of end vertices of heavier edges are more likely to be similar.}

\subsection{Low-Pass Graph Filter}\label{sec:filter}

The process of frequency filtering transforms and input signal into a linear combination of complex exponentials. As a consequence, filtering amplifies or attenuates the contributions of some frequencies. Graph spectral filtering can be directly generalized as

\begin{equation}\label{eq:filtro}
    f_{out}(\lambda_l) = f_{in}(\lambda_l)h(\lambda_l)
\end{equation}

\noindent where $h(\bullet)$ is the filter transfer function and  $f_{in}$ is the Fourier transform of an input signal function.  Several well-known continuous filtering techniques can be implemented as discrete, considering filtering functions that satisfy Equation \eqref{eq:filtro}, such as Gaussian smoothing, bilateral filtering and non-local means filtering \citep{Buades2005a}.

A filter is said to be low-pass if it does not significantly affect the frequency content of low-frequency signals but attenuate the magnitude of high-frequency signals. An ideal low-pass filter keeps the magnitude of the spectrum at low frequencies unchanged and attenuates it at high frequencies. The frequency response of these filters is defined as

\begin{equation}
    h(\lambda_l)= \alpha_l = 
    \begin{cases}
      1, & \lambda_l < \lambda_{cut} \\
      0, & \lambda_l \geq \lambda_{cut}
    \end{cases}
  \end{equation}
  
Sandryhaila and Moura \cite{Sandryhaila2014} demonstrated that the design of these filters is a linear problem and the construction of a filter with frequency response $h(\lambda_l) = \alpha_l$ corresponds to solving a system of $n_l$ non-linear equations:

\begin{align} \label{lowpass}
  h_0 + h_1\lambda_0 + ... + h_{d_p}\lambda_0^{d_p}=& \,\alpha_0, \\ 
  h_0 + h_1\lambda_1 + ... + h_{d_p}\lambda_1^{d_p}=& \,\alpha_1, \\ 
  \vdots\\
  h_0 + h_1\lambda_{n_l-1} + ... + h_{d_p}\lambda_{n_l-1}^{d_p}=& \,\alpha_{n_l-1}
\end{align}

\noindent where $d_p$ is the degree of the polynomial. We can find an approximate solution by, for example, the least squares method, to get around the over-determination of the system when $n_l \geq {d_p} + 1$. \mariah{The complexity of the least squared method for determining a triangular $n_l$-dimensional matrix inverse is $O(n_l^2)$.}

\section{Related Works} \label{sec:rw}

An important task in data mining is finding instances with unexpected behavior, which are more likely to be anomalous observations. Although several techniques have been developed over the last years to identify anomalies in data \citep{li2021clustering}, there are few techniques capable of efficiently dealing with graph-structured data. Graph structured data have complex correlations and require techniques capable of analyzing not only the data itself but the relations between the elements. This paper is particularly interested in graphs with node attributes to find community outliers. A community outlier is a node whose attribute values deviate significantly from its community members.

In a recent survey on graph-based anomaly detection, Akoglu et al. \cite{Akoglu2015GraphSurvey} pointed out only two community-based methods for detecting anomalies in graphs with attributes. The first, introduced in \citep{GaoOn}, is a probabilistic model that considers both relational and raw data information to find more meaningful outliers. On the one hand, the information regarding the relation is obtained from the topological structure of the network and describes the relationships that exist between instances of the dataset. On the other hand, the network node attributes store the data. According to the authors, the algorithm, called community outlier detection algorithm (CODA), can identify meaningful community outliers. The second method is a node outlier ranking technique in attributed graphs, called GOutRank, developed by M{\"u}ller et al.  \cite{Muller2013}. GOutRank ranks the graph nodes according to their degree of deviation in both graph-relational data and node attribute properties. 

In a particular case of attributed graphs, where the attributes of the nodes can be interpreted as a signal indexed by the graph, techniques that extend signal processing concepts to the graph domain can be used to identify anomalies. For example, attributed graphs can represent wireless sensor networks, where each (sensor) node stores the value measured by the corresponding sensor in an instant of time. In this case, to identify an anomaly, it is necessary to consider the geographical proximity between the sensors and the values sensed by them. A sensor measurement significantly different from neighboring sensors' represents a potential anomaly. For this type of application, classical signal processing techniques, such as filters, can be extended to the graph domain.

Several studies use graph-based filters for data compression, data recovery, classification, noise removal, signal recovering, among others. However, to the best of our knowledge, few studies have adopted graph-based filtering to detect anomalies. Sandryhaila and Moura \cite{Sandryhaila2014}, for example, introduced the concept of total variation to frequency sorting. The authors developed a filter using the proposed ordering method to identify malfunctions in sensor networks by extracting high-frequency components.  Egilmez and Ortega \cite{Qualcomm2014SpectralGraphs} introduced a spectral anomaly detection method that also uses a graph-based filter. They considered collective anomalies which occurred locally in time and space and applied the proposed method to sensor networks. 

Both \cite{Sandryhaila2014} and \cite{Qualcomm2014SpectralGraphs}  adopted spectral filters and presented experiments with sensor network data where the adjacency relationships may be sufficient to detect nodes whose attributes deviate from the attributes of other nodes from the same community. However, in applications where the adjacency relationships are not only based on the physical distance, identifying a community outlier may consider the network's community structure. Neither of the previous work takes such information into account in their analysis. 

In protein-protein interaction networks, for example, adjacency relationships define the biological strength of the interaction between a pair of proteins. Proteins, represented by the graph nodes, may have quantitative attributes that correspond to the expression level of the protein in the network. In these networks, a community can be interpreted as a group of functionally related proteins \citep{francisquini2021community} whose attributes have similar characteristics. For this type of application, analyzing only the adjacency affinities may not reveal relevant information. Then, the data analysis must explicitly consider the community structure. To the extent of our knowledge, there is no graph-based filter method to detect anomalies that also takes the community structure into account, as proposed here.

\section{Anomaly Detection} \label{sec:ad}

The anomaly studied in this paper is defined as a vertex $v_i$ whose signal $f(v_i)$ is far from the expected standard found in the community to where  $v_i$ belongs. In this case, most of the signal’s energy is concentrated in the low-frequency signals of the Fourier transform. A node with a signal value that considerably differs from the neighborhood average will probably be anomalous. 

This type of anomaly is observed, for example, in temperature sensor networks, in which neighboring sensors are expected to have close measured values. In this case, a potential anomaly is a sensor measurement that deviates significantly from the values sensed by neighboring nodes, such as a sensor failure. In protein-protein interaction networks, expression levels of neighboring nodes that represent proteins belonging to the same biological process usually vary proportionately. For example, if the expression level of a protein increases, other proteins that belong to the same biological process usually have their expression level increased or decreased accordingly. In this case, where proteins from the same biological process generally belong to the same community, a potential anomaly is a node with a variation of expression level significantly different from the nodes in the same community, e.g., a protein involved in cancerous processes.

The reasoning behind the proposed strategy is that the attenuation of the magnitude of the signal spectrum at high frequencies can correct the abrupt variations that define the studied anomalous behavior. Thus, the  anomaly detection method introduced in this paper, \textit{SpecF}, relies on the idea that vertices whose signal value needed a substantial correction after the filtering process are more likely to be anomalous. For this, \textit{SpecF} uses a low-pass filter to attenuate the magnitude of the high frequencies of the spectrum $\hat{\mathbf{f}}$ of a signal $\mathbf{f}$. The inverse Fourier transform, defined in Equation \eqref{eq:inversefourier}, is applied to the filtered spectrum to obtain a filtered signal $\mathbf{f'}$. The filtered signal $\mathbf{f'}$ is compared to the original signal $\mathbf{f}$ to determine to which vertices the signal value has been severely attenuated, to obtain the set of potentially anomalous vertices. \mariah{Algorithm~\ref{alg:specf} presents a basic pseudocode of \emph{SpecF}, whose main steps will be thoroughly described in the next sections. Besides the  graph and its signals, the input data required by this algorithm are a matrix representing $G$, which can be the adjacency matrix, and a partition $\mathcal{C}=\{C_1, C_2, \ldots, C_{|\mathcal{C}|}\}$ representing the set of communities of $G$.}

\begin{algorithm}[!h]
\small
\mariah{\KwData{A graph $G$,  a matrix $M_G$ representing $G$, signal $B$, partition $\mathcal{C}$ }
\KwResult{A list with the anomalous nodes $PAN$}
$L:=D-M_G$\\
Calculate matrix $U$, the set of eigenvectors of $L$\\
Calculate the Fourier transform using Equation~\eqref{eq:fourier} to obtain the spectrum $\hat{B}$ of $B$:  $\hat{B} = U^TB$ \\ 
$B' \leftarrow$ Low-pass filter ($G, L, U, \hat{B}$) -- Algorithm~\ref{alg:algoritmo4} discussed in Section~\ref{sec:lowpass}\\
$PAN\leftarrow$ Potentially anomalous nodes($G, B,B', \mathcal{C}$) -- Algorithm~\ref{alg:algoritmo5}  discussed in Section~\ref{subsec:anomalous}\\
}\caption{\mariah{\emph{SpecF}}}
\label{alg:specf}
\end{algorithm}

To evaluate the proposed strategy, this paper also introduces an approach to generate synthetic anomalous signals similar to the signals considered in the hypothesis. The next section discusses the details of  the proposed anomaly detection algorithm.

\subsection{Low-pass Filter and the Cut-off Frequency}\label{sec:lowpass}

In GSP theory, the first $k$ eigenvalues of $L$ correspond to the $k$-lowest frequencies in the spectrum of a signal. Low frequencies carry the information of signals that vary slightly across the nodes of the network. Moreover, the proposed strategy focuses on a signal whose intra-community variation is expected to be low. Therefore, the introduced method considers the number of communities $k$ in the network as the cut-off frequency, so that $\lambda_{cut} = \lambda_k$. A low-pass filter, as defined in Section \ref{sec:filter}, attenuates the magnitude of high-frequency signals and keeps low frequencies unchanged. In this case, a high-frequency is defined as a frequency $\lambda_l$ higher than the cut-off frequency $\lambda_{cut}$, that is, $\lambda_l > \lambda_{cut}$.

In cases where the expected partition is known beforehand, the choice of $\lambda_k$ is trivial, since $k$ is the number of communities. On the other hand, when the number of communities is unknown,  two approaches to estimate the number of communities in the network can be used. The first approach consists of applying a community detection algorithm to the network to estimate the number of communities $k$ -- algorithms that do not require the number of communities to be informed \textit{a priori}. The second approach comes from graph theory and consists of finding the value of $k$ by analyzing the eigenvalues of $L$ and choosing $k$ so that $\lambda_1, ..., \lambda_k$ are very small, but $\lambda_{k+1}$ is relatively large. In spectral graph theory, when a graph has $k$ completely disconnected components, the first $k$ eigenvalues of $L$ will have the value 0 and then there is a gap to the eigenvalue in position $k+1$ \citep{VonLuxburg2007AClustering}. Algorithm~\ref{alg:algoritmo4} shows a pseudocode to determine the low-pass filter for the proposed anomaly detection algorithm. \mariah{The complexity of Algorithm~\ref{alg:algoritmo4} is dominated by the least square method to the system \eqref{lowpass}, which has been discussed earlier, $O(n^2)$.}

  

\begin{algorithm}[!h]
\small
\mariah{\KwData{A graph $G$, the Laplacian matrix $L$, matrix $U$, spectrum $\hat{B}$}
\KwResult{Filtered signal $B'$}
Estimate the number of communities $k$ of the input graph $G$ by analysing the eigenvalues of $L$ (second approach)\\
Find the $h_l$ values by solving the system~\eqref{lowpass}, where $n_l$ and $d_p$ are $n$\\
 Define a diagonal matrix $\mathcal{F}$ where its $i$-th diagonal element  is the approximate $\alpha_i$ -- according to system~\eqref{lowpass}\\
The $\hat{B}$ spectrum is submitted to the proposed low-pass filter to obtain a filtered spectrum $\hat{B}'$: $\hat{B}' = \mathcal{F}\hat{B}$ \\
 Then, the inverse Fourier transform is applied in $\hat{B}'$ to obtain a filtered signal $B'$: ${B}' = U  \hat{B}'$ \\}
\caption{\mariah{Low-pass filter}}
\label{alg:algoritmo4}
\end{algorithm}

\begin{figure}[htbp]
\center

\input{tikz/anomalous_ordered2.tex}

\caption{Example of a filtered signal $B'$.}
\label{fig:anomalia2}
\end{figure}
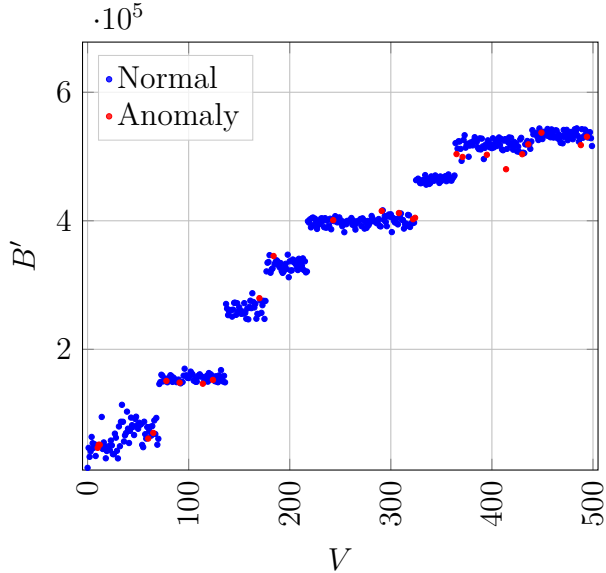

Figure \ref{fig:anomalia2} plots the filtered signal $B'$ of the signal presented in Figure~\ref{fig:anomalia1}. By comparing the signal $B'$ to the signal $B$, it is possible to observe that the groups of vertices are more cohesive \textcolor{black}{ in signal $B'$}, demonstrating that the filter brought the signal values even closer to nodes from the same community. In addition, the anomalies, which previously diverged from other nodes in the same community, are now within the expected standard.

\subsection{Potentially Anomalous Nodes}\label{subsec:anomalous}

Figure \ref{fig:anomalia2} shows the ability of the \textit{SpecF} to correct anomalies and normalize the values of a signal according to the network's community structure. The filtered signal $B'$ is contrasted to the original anomalous signal $B$ to identify in which nodes the correction of the signal was more significant. The intuition behind this idea is that, if anomalous nodes differ from what is expected for their community, the normalization applied to the signal considering the filter will be more intense in these nodes. Thus, when identifying the anomalous nodes, a set of nodes with the greatest anomalous potential is also detected. For such, let $Y = \abs{B - B'}$ be a signal, and its $i$-th element be referred to as $y_i$, corresponding to the difference signal at vertex $v_i$. In general, vertices $v_i$ with a high $y_i$ value are more likely to be anomalous and the vector $Y$ is deemed an abnormality quantifier. 

To classify the nodes as anomalous or normal in a binary way, \emph{SpecF} applies to $Y$ a threshold to distinguish which $y_i$ values are considered normal and which are identified as abnormal observations. The threshold values employed in \textit{SpecF} are regarded for each community, taking the mean and standard deviation of $Y$ into account, as presented in Equation~\eqref{eq:threshold}. 

\begin{equation}
    TD(Y, C_k) = \ddfrac{mean(Y, C_k) + 2std(Y, C_k)}{max(Y, C_k)}
    \label{eq:threshold}
\end{equation}

\noindent  $mean(Y, C_k)$, $std(Y, C_k)$ and $max(Y, C_k)$ are, respectively, the mean, standard deviation and  maximum of the values of $y_i$'s that represent the nodes of community $C_k$. Algorithm~\ref{alg:algoritmo5} presents a pseudocode of the strategy that defines the potentially anomalous nodes.  In this algorithm,  let $c_{v_i}$ be the community $C_k$ where vertex $v_i$ belongs to. \mariah{The complexity of Algorithm~\ref{alg:algoritmo5} is $O(n)$.}


\begin{algorithm}[!h]
\small
\mariah{\KwData{A graph $G$,  signal $B$, filtered signal $B'$, partition $\mathcal{C}$}
\KwResult{A list $PAN$ with the potentially anomalous nodes}
$Y = \abs{B - B'}$\\
$PAN=\emptyset$\\
Insert in $PAN$ every node $v_i\in V(G)$ whose $TD(Y,c_{v_i})$ is lower than $y_i$}
\caption{\mariah{Potentially anomalous nodes}}
\label{alg:algoritmo5}
\end{algorithm}

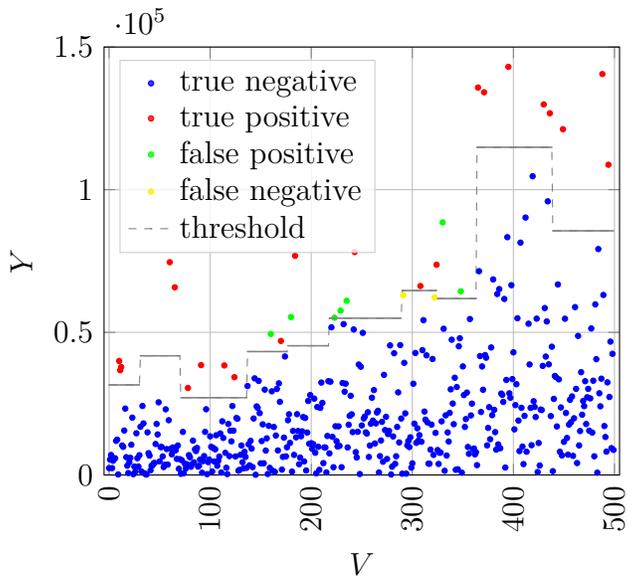
\begin{figure}[htbp]
\center

\input{tikz/anomalous_type.tex}

\caption{Example of a signal $Y$ and the threshold.}
\label{fig:anomalia3}
\end{figure}

Figure \ref{fig:anomalia3} illustrates the relation between $y_i$ values and the vertices of a network. Moreover, it presents the threshold values $TD(Y, C_k)$ -- dotted curve --  that separate  anomalous nodes from  normal nodes. It is possible to notice that the vast majority of nodes above the red curve are true positives and, therefore, are in the set of anomalous nodes defined by the generator. On the other hand, most of the nodes below the red curve are true negatives and, thus, are labeled normal. There are also false positives and false negatives and they are usually close to the threshold curve.

\mariah{The computational complexity of \textit{SpecF}, described in Algorithm~\ref{alg:specf}, is $O(n^2)$ since to calculate the Fourier transform a matrix multiplication operation is required.}

\subsection{Expanded Adjacency Matrix}\label{expand_matrix}

\mariah{To embed to a matrix information about its community structure, this paper proposes the use of an expanded adjacency matrix to represent a graph $G$ -- one of the forms to define matrix $M_G$. } 

Let $W \in \mathbbm{R}^{n\times n}$ be the so-called expanded matrix that incorporates the community structure of a graph to represent the pairwise relationship between vertices $v_i$ and $v_j \in V$, referred to as $w_{ij}$. The values $w_{ij}$ are defined in Equation \eqref{case:case1},  where $c_{v_i}$ is the community where vertex $v_i$ belongs to.

\begin{equation}\label{case:case1}
w_{ij}=
\begin{cases}
    5,& \text{if $v_i$ and $v_j$ are neighbors and $c_{v_i}$ = $c_{v_j}$}\\
    3,& \text{if $v_i$ and $v_j$ are neighbors and  $c_{v_i} \neq c_{v_j}$}\\
    1,& \text{if $v_i$ and $v_j$ are not neighbors and  $c_{v_i} = c_{v_j}$}\\
    0,              & \text{otherwise}
\end{cases}
\end{equation}

According to the definition of the expanded matrix, if vertices $v_i$ and $v_j$ are adjacent and in the same community,  $w_{ij}$ will receive the highest weight, the value 5. If they are adjacent but belong to distinct communities, $w_{ij}$ will receive the intermediary value 3.  If the two vertices are not adjacent but are in the same community, $w_{ij}$ will receive the value of 1. As a consequence, an explicit affinity between these vertices is defined in such a matrix. Other edges assume null weight and are disregarded.

\subsection{Attributed Networks Generator} \label{cap:generator}
\mariah{This section introduces the methodology to generate anomalous and normal signals in networks with community structure.  }

\subsubsection{Normal Signal Generator} \label{cap:normalgenerator}

Let $G^c$ be a graph whose nodes $v^c_i$ represent communities $C_i$ of $G$ and the edge weights $w^c_{ij}$ are defined as the number of edges between communities $C_i$ and $C_j$. To define a signal $S^c$, the nodes of $G^c$ are sorted according to the sum of the edges' weights $w^c_{ij}$ and $s_i^c$ is then defined as  


\begin{equation} \label{eq:signal_sc}
s_i^c = \sum_{\forall v^c_j \in \mathcal{N}_{i}^c} w^c_{ij} \times (i+1)
\end{equation}

\noindent where $\mathcal{N}_{i}^c$ is the set of nodes adjacent to $v_i^c$.

To properly evaluate the anomaly detection strategy proposed in this paper, we developed an artificial signal generator. The generator produces a synthetic signal $S$ similar to the signal observed in the investigated applications. As a result, it creates signals whose values for nodes of the same community are similar. 

Algorithm \ref{alg:algoritmo2} describes the process employed by the generator to obtain the signal $S$ from a graph $G$ and a given signal $S^c$, with elements defined by Equation \eqref{eq:signal_sc}. First, all nodes are marked and an auxiliary $n$-dimensional vector $S^x$ is initialized as empty. For every community $C_k$ of $G$, the algorithm assigns the value of $s_k^c$ to the position of $S^x$ that corresponds to the highest degree node in  the community $C_k$. In the case of a tie, a node is randomly selected among the highest degree nodes.  These nodes are regarded as community heads. Then, the algorithm starts a propagation process from the community heads. Each node propagates a percentage of its value to its neighbors. This percentage is lower (10\%) if the nodes involved belong to different communities, and higher (at least 25\%) if they are in the same community. For nodes from the same community, this percentage also depends on their degree, so that nodes with higher degrees have a greater influence on lower degree nodes. After the propagation process, every node that propagated values have their value reduced by 5\%. This process is repeated until all nodes have propagated a number of their signal values. The last step consists of a normalization process to define the signal of each node as that considers the weight of the edge between neighbor nodes to carry out a weighted average.

\begin{algorithm}[!ht]
\small
\KwData{A graph $G$, a signal $S^c$, a partition $\mathcal{C}$ }
\KwResult{A signal $S$}

Mark all nodes\;
$S^x \leftarrow \emptyset$\;

\ForAll{ community $C_k\in \mathcal{C}$}{
    $v_i \leftarrow$ the highest degree node of $C_k$\;
    $s^x_i \leftarrow s^c_k$\;
    Unmark $v_i$\;
}


Create a list $F$ with all unmarked nodes sorted by index values\;
\While{there is a node in $F$}{ 
    \textcolor{black}{Select the first node $v_i$ from $F$}\;
    \ForAll{$v_j\in \mathcal{N}_i$ }{
        $t \leftarrow 0.1$\;
        \If{$v_i$ and $v_j$ are in the same community}{
            $mt \leftarrow  \ddfrac{degree(v_i)}{degree(v_j) + degree(v_i)}$\;
            $t \leftarrow max(0.25, mt)$\;
        }
        $s^x_j \leftarrow s^x_j + (s^x_i \times t$)\;
        $s^x_i \leftarrow s^x_i \times 0.95$\;
        \If{$v_j$ is marked}{
            Unmark $v_j$ and include $v_j$ at the end of $F$\;
        }
    }
    Remove $v_i$ from $F$\; 
}
\ForAll{$v_i\in G$}{
    $s_i \leftarrow   \frac{\sum_{j=1}^n w_{ij}\times s_j^x}{\sum_{j=1}^n w_{ij}}$\;
}
\caption{\textsc{Normal Signal Generator}}
\label{alg:algoritmo2}
\end{algorithm}

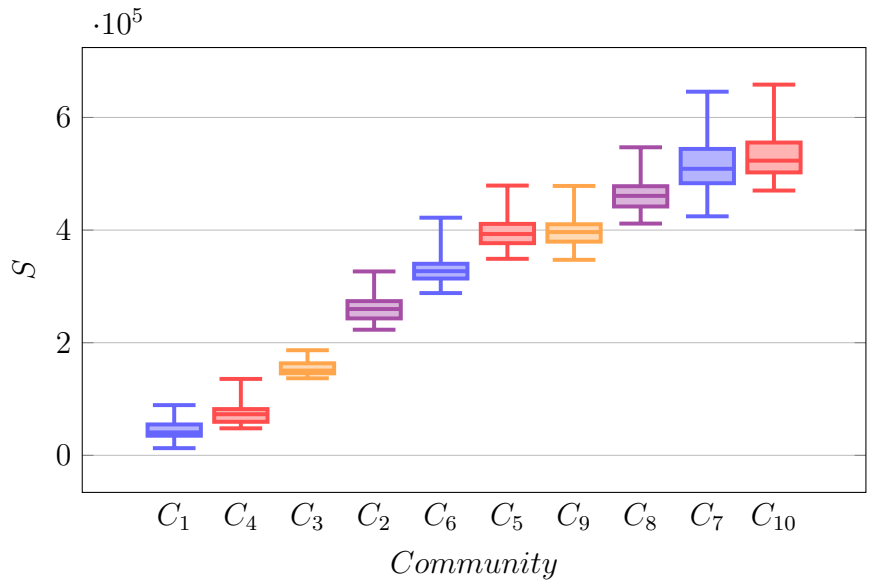
\begin{figure}[h!t]
\center

\input{tikz/box_GLOBAL.tex}

\caption{Box-plots of $S$ for each community of $G$.}
\label{fig:gerador2}
\end{figure}

Figure \ref{fig:gerador2} presents box-plots of the values of the $S$ signal, obtained through Algorithm \ref{alg:algoritmo2} applied to a 500-node LFR network \citep{Lancichinetti2009} with the parameters defined in Table \ref{tab:parameters} and low overlapping between communities (mixture degree 0.1).  The network has 10 planted communities. Section~\ref{subsec:artificial} describes the parameters and the software to generate the networks. We show a box-plot for each of the expected communities, sorted by increasing order of intra-community average signal.
Figure \ref{fig:gerador2} shows that, although the average value of signal $S$ in each community is different, vertices in the same community have similar values.


\subsubsection{Anomalous Signal Generator}\label{sec:anomalygerador}

In addition to the normal signal generator, a strategy to create anomalous signals in an attributed network is introduced in this paper. 

For such, consider a signal $S$ of a graph $G$. We generate an anomalous signal $B$ from signal $S$ by increasing the value of $s_{i}$ in some vertices of $G$. This process is detailed in Algorithm \ref{alg:algoritmo3}, which randomly selects a set of nodes to corrupt. The anomaly intensity $\theta$ defines how much greater the value of an anomalous node will be when compared to the rest of the community. An anomaly intensity value of $0.1$, for example, means that an anomalous node will have a signal value between 5\% and 10\% higher than the largest signal value of that community.

\begin{algorithm}[!h]
\small
\KwData{A graph $G$, a normal signal $S$, percentage of anomalies $AN$, anomaly intensity $\theta$}
\KwResult{An anomalous signal $B$}
$B \leftarrow$ Create a copy of $S$\;
$P \leftarrow$ Randomly select a set with $AN\%$ of distinct vertices of $G$\;
\ForAll{vertex $v_i$ in $P$}{ 
    $max\leftarrow$ Highest value of $S$ among nodes in $c_{v_i}$\;
    $tax \leftarrow$ Random value between $\frac{1}{2}\theta$ and $\theta$\;
    $b_i$ = $max \times (1 + tax)$\;
}
\caption{\textsc{Anomalous Signal Generator}}
\label{alg:algoritmo3}
\end{algorithm}

\begin{figure}[htbp]
\center

\input{tikz/anomalous_ordered.tex}

\caption{Example of an anomalous signal $B$.}
\label{fig:anomalia1}
\end{figure}
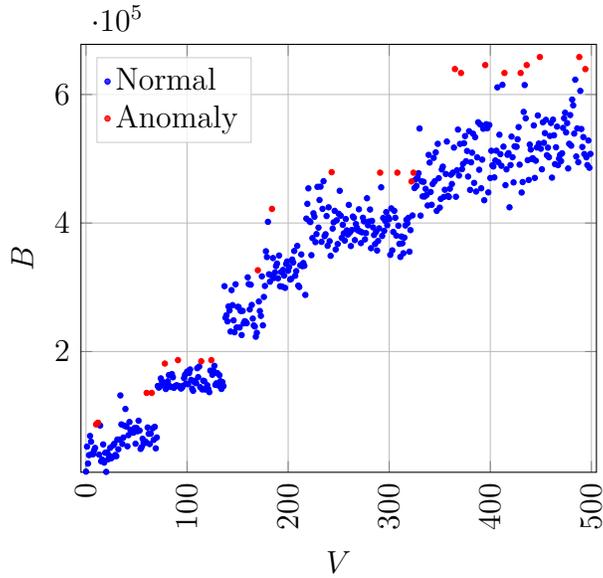

Figure \ref{fig:anomalia1} illustrates the values of an anomalous signal $B$ at each node. Anomalies are highlighted  (when `Anomaly' is 1). Moreover, the vertices are represented in the x-axis, sorted according to the intra-community average signal, as in Figure \ref{fig:gerador2}. The vertices within communities are ordered by index. It is possible to see that, when sorted by the community average, the anomalous nodes become more evident. However, they are within the mean and standard deviation values when considering the complete signal, which makes them difficult to detect using techniques that do not consider the network community structure.


The next section presents the computational experiments performed to evaluate the efficacy of \textit{SpecF}.

\section{Computational Experiments}\label{sec:ce}

This section presents five experiments carried out to attest the efficiency of the anomaly detection method proposed in this paper. The first three experiments were carried out with artificial networks, whereas the forth and fifth consist of tests performed with real-world datasets.  The first experiment assesses the behavior of \textit{SpecF} by varying the number of anomalies present in the network. The second experiment analyzes \textit{SpecF} when faced with variations in the intensity of the anomalies. The third experiment evaluates the effectiveness of the strategy in multiple executions with varied parameter values. \mariah{Also an experiment with labeled data, the forth experiment employs two publicly available IoT datasets, where a comparative analysis with state-of-the-art algorithms is performed.} The last experiment presents a thorough analysis of an unlabeled COVID-19 dataset. 

Before going into detail about the experiments with artificial networks, the generated networks and employed evaluation metrics are discussed.

\subsection{Artificial Networks}\label{subsec:artificial}

A set of artificial networks was generated using the software introduced in \citep{Lancichinetti2009}, referred to as LFR networks. By using this software, a set of undirected and unweighted benchmark graphs, with heterogeneous distributions of node degree and community sizes is created. The nodes of the generated LFR networks have an average degree $d_G$ of 10 and a maximum degree $max$ $d_G$ of 50. The parameters related to the exponent of the distribution of degrees (neg. exp. $d_G$) and community vertex count (neg. exp. $\mid C\mid$) are  2 and 1, respectively. The mixing parameter ($\mu$) was set to be valued between 0.1 and 0.8, with a step size of 0.1. The mixture degree of the communities reflects how well separated the communities are since $\mu$ specifies the amount of inter-community edges. Therefore, low values for the mixture parameter produce networks with a more evident division in communities.  
The strategy introduced to generate normal and anomalous signals discussed in  Section~\ref{cap:generator} was applied to the LFR networks.
  More details regarding the anomaly intensity ($\theta$) and percentage of anomalies (\textit{AN}) values are approached in the experiments. Table \ref{tab:parameters} summarizes the LFR and signal parameters used to generate the networks.

\begin{table}[t]
\begin{center}

\caption{Parameters employed to generate the LFR  networks and the normal/anomalous signal.}
\label{tab:parameters}
\begin{tabular}{@{} *5l @{}}    \toprule
Type &{Parameter} & {Values} \\\midrule
\multirow{7}{*}{LFR} &$n$  & 500 and 1000\\
 &$\mu$ (mixture parameter) & 0.1, 0.2, \ldots, 0.8\\
 &av. $d_G$ & 10 \\
 &max $d_G$ & 50 \\
 &neg. exp. $d_G$ & 2 \\
 &neg. exp. $\mid C\mid$ & 1\\
 &min/max vertex count in communities & 20/100 \\
 \hline
 Signal&$\theta$, \textit{AN} (\%)& 1, 5, 10, 15, 20\\\bottomrule
 \hline
\end{tabular}

\end{center}
\end{table}

\begin{figure}
	\centering
		\includegraphics[scale=.12]{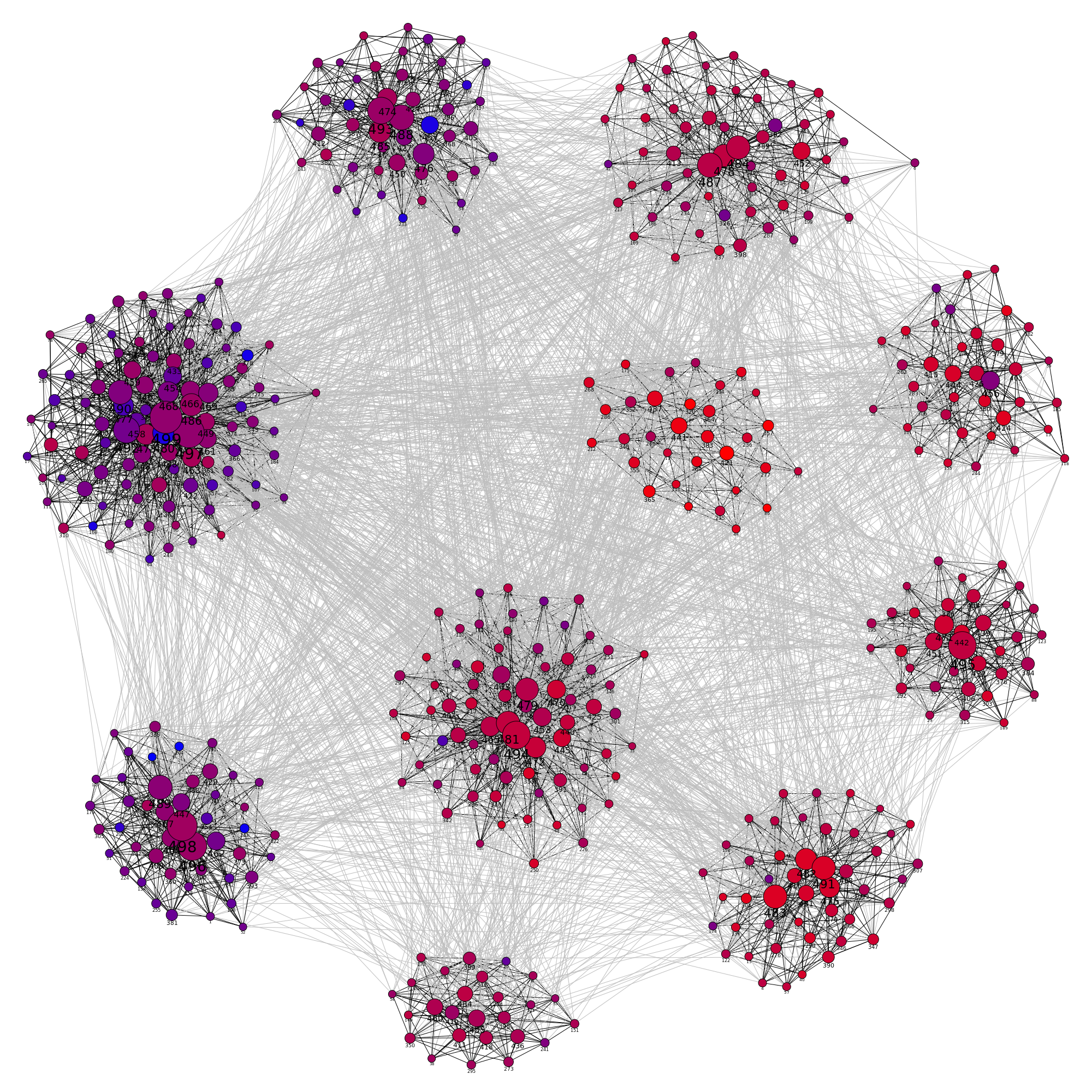}
	\caption{Example of an LFR network with anomalies.}
	\label{fig:lfr_network}
\end{figure}

Figure \ref{fig:lfr_network} illustrates an LFR network with 1000 vertices colored according to the signal values. The bluer a node, the higher its signal value. On the other hand, the redder the nodes, the lower their signal values. Larger nodes are those with a higher number of neighbors. The vertices are separated into different groups that represent the communities to which they belong. One may observe that intra-community vertices have similar colors, for example. However, there are some anomalous vertices, like the blue ones, that subtly clash with the standard of the vertices of their community.

\subsection{Evaluation Metrics}

Classic measures were used to evaluate and interpret the results obtained by \textit{SpecF}. 

The Receiver Operating Characteristic curve, or ROC curve, summarizes the trade-off between the false positive rate and true positive rate (also known as recall). The ROC curve allows the comparison of different models directly, and the area under the curve (AUC) can be used as a model quality quantifier. On the one hand, a random classifier, for example, could be represented by a diagonal curve that has an area of 0.5, starting at the bottom left and ending at the top right of the ROC space. On the other, a curve that starts in the lower-left corner, moves up to the upper left corner, and then advances to the upper right corner, adding up to an area of 1 would represent an ideal classifier.


In cases of binary classification problems with a skewed distribution of numbers of observations per class, Saito and Rehmsmeier \cite{Saito2015TheDatasets} point out the precision-recall curve as a more informative metric. Since anomaly detection is often characterized by a skewed distribution (there are far more normal cases than anomalies), the precision-recall curve in the evaluation of the anomaly detection performance is also presented here. The precision-recall curve is similar to the ROC curve, but it compares precision with recall for different thresholds. Precision corresponds to the proportion of true positives concerning the sum of true positives and false positives. A random classifier, in this case, is represented by a horizontal line with a value proportional to the number of positive cases in the data set. An efficient classifier, on the other hand, would be represented by a line that approaches the upper right point of the plot. Again, the Area Under the Precision-Recall Curve (AUC-PR) can be used as a quantifier of the classification model's ability, also called Average Precision (AP) \citep{Saito2015TheDatasets}. 

\subsection{Experiment I}\label{subsec:experiment1}

The first experiment seeks to evaluate the performance of \textit{SpecF} by varying the number of anomalies and keeping the intensity $\theta$ at 5\%. \mariah{To assess the robustness of the method, this experiment employs a total of 250 networks generated by considering the following methodology:

\begin{itemize}
    \item Generate five different LFR networks with $\mu=0.1$, 500 nodes and the remaining LFR parameters at the values presented in Table~\ref{tab:parameters};
    \item Apply the normal signal generator algorithm (Algorithm ~\ref{alg:specf}) once to each of the five networks;
    \item Apply the anomalous signal generator algorithm (Algorithm~~\ref{alg:algoritmo2}) to each network ten times for each of the \textit{AN} values presented in Table~\ref{tab:parameters} and fixing $\theta$ at value 5\%. Therefore, this step produces 50 different networks (one for each execution of the anomalous signal generator) for each \textit{AN} value, totaling 250 networks.
\end{itemize}
}

Table \ref{tab:experiment1} presents the mean and standard deviation of the AUC-ROC and AP values considering the 50 networks per \textit{AN}, when a standard adjacency matrix is used in \textit{SpecF}. These results show that the performance of \textit{SpecF} improves as $AN$ increases. Similarly, Table \ref{tab:experiment1b} presents the results of the same experiment, but using the expanded adjacency matrix $W$ instead of the adjacency matrix $A$ to calculate the Fourier transform defined in Section \ref{sec:fourier}.  A comparison of the results with both matrices reveals that in all cases the anomaly detection strategy performed much better using the expanded adjacency matrix $W$\mariah{, for all $AN$ values and both AUC-ROC and AP metrics}. 


\begin{table}[t]
\begin{center}
\caption{Results of experiment I using the  standard  adjacency matrix}
\label{tab:experiment1}
\begin{tabular}{@{} *5l @{}}    \toprule
$AN$ & AUC-ROC & AP \\\midrule
$1\%$  &  0.899 $\pm$ 0.042 & 0.511 $\pm$ 0.099 \\
$5\%$  &  0.936 $\pm$ 0.029 & 0.639 $\pm$ 0.098 \\
$10\%$  &  0.928 $\pm$ 0.023 & 0.704 $\pm$ 0.067 \\
$15\%$  &  0.953 $\pm$ 0.017 & 0.808 $\pm$ 0.050 \\
$20\%$  &  0.949 $\pm$ 0.023 & 0.793 $\pm$ 0.066 \\

 \hline
\end{tabular}
\end{center}
\end{table}



\begin{table}[t]
\begin{center}
\caption{Results of experiment I using the expanded adjacency matrix}
\label{tab:experiment1b}
\begin{tabular}{@{} *5l @{}}    \toprule
$AN$ & AUC-ROC & {AP} \\\midrule
$1\%$  &  0.931 $\pm$ 0.016 & 0.513 $\pm$ 0.050 \\
$5\%$  &  0.958 $\pm$ 0.014 & 0.682 $\pm$ 0.068 \\
$10\%$  &  0.975 $\pm$ 0.008 & 0.785 $\pm$ 0.053 \\
$15\%$  &  0.969 $\pm$ 0.015 & 0.810 $\pm$ 0.037 \\
$20\%$  &  0.978 $\pm$ 0.008 & 0.826 $\pm$ 0.048 \\

 \hline
\end{tabular}
\end{center}
\end{table}



\subsection{Experiment II}\label{subsec:experiment2}

The second experiment compares the accuracy of \textit{SpecF} considering different anomaly intensities and keeping $AN$ fixed at 5\%. \mariah{This experiment was carried out using the same networks generated in the second step of the methodology presented in the earlier section. Therefore, for this experiment, the third step of the methodology, the one that produces the anomalous signals, consists in following the procedure:

\begin{itemize}
    \item Apply the anomalous signal generator algorithm (Algorithm~\ref{alg:algoritmo2}) to each network ten times for each of the $\theta$ values presented in Table~\ref{tab:parameters} and fixing \textit{AN} at 5\%.
\end{itemize}
}
 Table \ref{tab:experiment2} reports the results of \textit{SpecF} when the standard adjacency matrix is used. In cases where the anomaly is extremely hard to identify, as in cases where the value of the signal at the anomalous nodes are only 1\% greater than the maximum signal in their community, the accuracy of the model is poor. In the case with anomaly intensity $\theta = 1\%$, for example, the mean AP was approximately 0.43, which is very close to the values obtained by a random classifier. Table \ref{tab:experiment2b} presents the results of this experiment using the expanded adjacency matrix $W$ instead. Again, the use of the matrix $W$ improves the results in anomaly detection in all scenarios\mariah{, despite the anomaly intensity value and performance metric considered}.

\begin{table}[t]
\begin{center}
\caption{Results of experiment II using the standard adjacency matrix}
\label{tab:experiment2}
\begin{tabular}{@{} *5l @{}}    \toprule
\emph{$\theta$} & Mean AUC ROC & Mean AP \\\midrule
 $1\%$ & 0.925 $\pm$ 0.057 & 0.427 $\pm$ 0.169 \\
 $5\%$  & 0.923 $\pm$ 0.035 & 0.628 $\pm$ 0.080 \\
 $10\%$  & 0.900 $\pm$ 0.020 & 0.681 $\pm$ 0.023 \\
 $15\%$  & 0.868 $\pm$ 0.016 & 0.673 $\pm$ 0.032 \\
 $20\%$  & 0.857 $\pm$ 0.021 & 0.681 $\pm$ 0.024 \\

 \hline
\end{tabular}
\end{center}
\end{table}

\begin{table}[t]
\begin{center}

\caption{Results of experiment II using the expanded adjacency matrix}
\label{tab:experiment2b}
\begin{tabular}{@{} *5l @{}}    \toprule
\emph{$\theta$} & Mean AUC ROC & Mean AP \\\midrule
 $1\%$ & 0.957 $\pm$ 0.030 & 0.431 $\pm$ 0.179 \\
 $5\%$  & 0.946 $\pm$ 0.017 & 0.648 $\pm$ 0.074 \\
 $10\%$  & 0.939 $\pm$ 0.015 & 0.716 $\pm$ 0.037 \\
 $15\%$  & 0.919 $\pm$ 0.017 & 0.730 $\pm$ 0.039 \\
 $20\%$  & 0.887 $\pm$ 0.014 & 0.697 $\pm$ 0.018 \\

 \hline
\end{tabular}
\end{center}

\end{table}


 
\subsection{Experiment III}

This experiment evaluates the performance of \textit{SpecF} in a considerably larger set of networks. \mariah{The methodology to generate the set of 7200 networks had the following steps:

\begin{itemize}
    \item Generate five different networks for each pair $(\mu,n)$ of values presented in Table~\ref{tab:parameters}, totaling 80 different networks;
    \item Apply the normal signal generator (Algorithm~\ref{alg:algoritmo2}) once to each of the 80 networks (40 networks with 500 nodes and 40 networks with 1000 nodes);
    \item Apply the anomalous signal generator (Algorithm~\ref{alg:algoritmo3}) to each network ten times for each of the nine possible pairs $(\theta, AN)$, where  $\theta, AN \in \{1,5,10\}$, totaling 400 networks for each triple $(n,\theta, AN)$.
\end{itemize}
}


Tables  \ref{tab:experiment3.1} and  \ref{tab:experiment3.2} report the mean AUC-ROC  and AP for each triplet $(n,\theta, AN)$, considering \emph{SpecF} with the adjacency ($A$) and expanded ($W$) matrices. Therefore, each row of these tables corresponds to the average results of 400 different networks. It is possible to observe that, in  all cases, better results are obtained when the expanded adjacency matrix $W$ is used instead of the adjacency matrix $A$ in \textit{SpecF}.


\begin{table}[t]
\begin{center}
\caption{Results of experiment III varying different parameters for $n = 500$}
\label{tab:experiment3.1}
\begin{tabular}{@{} *6l @{}}   \toprule
\emph{$n$} & \emph{$AN$}  & \emph{$\theta$}  & $M_G$  & Mean AUC ROC  & Mean AP  \\\midrule
500 & 1\% & 1\% & $A$ & 0.756 $\pm$ 0.16 & 0.170 $\pm$ 0.15 \\
500 & 1\% & 1\% & $W$ & \textbf{0.956 $\pm$ 0.04} & \textbf{0.339 $\pm$ 0.18} \\\midrule
500 & 1\% & 5\% & $A$ & 0.797 $\pm$ 0.15 & 0.233 $\pm$ 0.18 \\
500 & 1\% & 5\% & $W$ & \textbf{0.971 $\pm$ 0.03} & \textbf{0.492 $\pm$ 0.21} \\\midrule
500 & 1\% & 10\% & $A$ & 0.847 $\pm$ 0.12 & 0.322 $\pm$ 0.21 \\
500 & 1\% & 10\% & $W$ & \textbf{0.990 $\pm$ 0.02} & \textbf{0.712 $\pm$ 0.22} \\\midrule
500 & 5\% & 1\% & $A$ & 0.740 $\pm$ 0.11 & 0.286 $\pm$ 0.13 \\
500 & 5\% & 1\% & $W$ & \textbf{0.947 $\pm$ 0.02} & \textbf{0.550 $\pm$ 0.11} \\\midrule
500 & 5\% & 5\% & $A$ & 0.780 $\pm$ 0.11 & 0.356 $\pm$ 0.14 \\
500 & 5\% & 5\% & $W$ & \textbf{0.970 $\pm$ 0.02} & \textbf{0.715 $\pm$ 0.12} \\\midrule
500 & 5\% & 10\% & $A$ & 0.836 $\pm$ 0.09 & 0.463 $\pm$ 0.16 \\
500 & 5\% & 10\% & $W$ & \textbf{0.985 $\pm$ 0.02} & \textbf{0.836 $\pm$ 0.13} \\\midrule
500 & 10\% & 1\% & $A$ & 0.714 $\pm$ 0.12 & 0.364 $\pm$ 0.13 \\
500 & 10\% & 1\% & $W$ & \textbf{0.931 $\pm$ 0.02} & \textbf{0.631 $\pm$ 0.09} \\\midrule
500 & 10\% & 5\% & $A$ & 0.763 $\pm$ 0.11 & 0.444 $\pm$ 0.14 \\
500 & 10\% & 5\% & $W$ & \textbf{0.958 $\pm$ 0.02} & \textbf{0.758 $\pm$ 0.09} \\\midrule
500 & 10\% & 10\% & $A$ & 0.815 $\pm$ 0.09 & 0.529 $\pm$ 0.13 \\
500 & 10\% & 10\% & $W$ & \textbf{0.975 $\pm$ 0.02} & \textbf{0.855 $\pm$ 0.10} \\
 \hline
\end{tabular}
\end{center}
\end{table}

\begin{table}[t]
\begin{center}

\caption{Results of experiment III varying different parameters for $n = 1000$}
\label{tab:experiment3.2}
\begin{tabular}{@{} *6l @{}}   \toprule
\emph{$n$} & \emph{$AN$}  & \emph{$\theta$}  & $M_G$  & Mean AUC ROC  & Mean AP  \\\midrule
1000 & 1\% & 1\% & $A$ & 0.747 $\pm$ 0.13 & 0.140 $\pm$ 0.11 \\
1000 & 1\% & 1\% & $W$ & \textbf{0.951 $\pm$ 0.03} & \textbf{0.292 $\pm$ 0.13} \\\midrule
1000 & 1\% & 5\% & $A$ & 0.792 $\pm$ 0.11 & 0.185 $\pm$ 0.14 \\
1000 & 1\% & 5\% & $W$ & \textbf{0.970 $\pm$ 0.03} & \textbf{0.444 $\pm$ 0.18} \\\midrule
1000 & 1\% & 10\% & $A$ & 0.848 $\pm$ 0.10 & 0.257 $\pm$ 0.16 \\
1000 & 1\% & 10\% & $W$ & \textbf{0.984 $\pm$ 0.02} & \textbf{0.614 $\pm$ 0.21} \\\midrule
1000 & 5\% & 1\% & $A$ & 0.739 $\pm$ 0.10 & 0.262 $\pm$ 0.12 \\
1000 & 5\% & 1\% & $W$ & \textbf{0.944 $\pm$ 0.02} & \textbf{0.543 $\pm$ 0.10} \\\midrule
1000 & 5\% & 5\% & $A$ & 0.782 $\pm$ 0.09 & 0.330 $\pm$ 0.13 \\
1000 & 5\% & 5\% & $W$ & \textbf{0.963 $\pm$ 0.02} & \textbf{0.672 $\pm$ 0.13} \\\midrule
1000 & 5\% & 10\% & $A$ & 0.823 $\pm$ 0.08 & 0.406 $\pm$ 0.14 \\
1000 & 5\% & 10\% & $W$ & \textbf{0.979 $\pm$ 0.02} & \textbf{0.793 $\pm$ 0.14} \\\midrule
1000 & 10\% & 1\% & $A$ & 0.721 $\pm$ 0.10 & 0.345 $\pm$ 0.11 \\
1000 & 10\% & 1\% & $W$ & \textbf{0.927 $\pm$ 0.02} & \textbf{0.623 $\pm$ 0.08} \\\midrule
1000 & 10\% & 5\% & $A$ & 0.757 $\pm$ 0.09 & 0.401 $\pm$ 0.12 \\
1000 & 10\% & 5\% & $W$ & \textbf{0.954 $\pm$ 0.02} & \textbf{0.742 $\pm$ 0.10} \\\midrule
1000 & 10\% & 10\% & $A$ & 0.809 $\pm$ 0.08 & 0.489 $\pm$ 0.13 \\
1000 & 10\% & 10\% & $W$ & \textbf{0.970 $\pm$ 0.02} & \textbf{0.825 $\pm$ 0.11} \\

 \hline
\end{tabular}

\end{center}
\end{table}



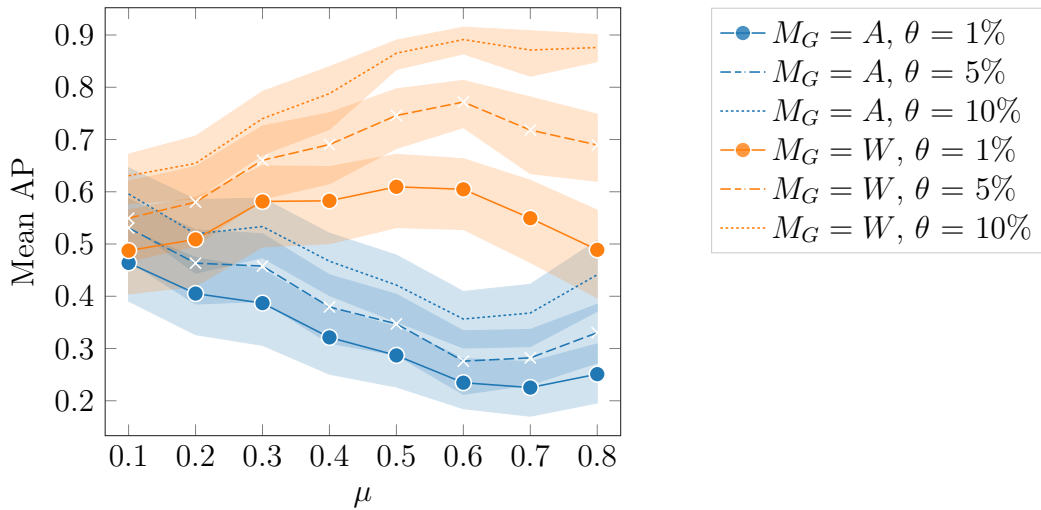
\begin{figure}[h!t]
\center

\input{tikz/output.tex}

\caption{Mean AP values for different values of $\mu$ and $\theta$.}
	\label{fig:mixture_types}
\end{figure}

Figure \ref{fig:mixture_types} illustrates the mean AP values for different values of $\mu$ and $\theta$ comparing \textit{SpecF} with $A$ and $W$ as input matrices. It is evident that, in all cases, \textit{SpecF} performed better when it adopted matrix $W$ instead of matrix $A$. Besides, the results are better for larger values of $\theta$, as the anomalies are more evident. The primary weakness of \textit{SpecF} considering both matrices is apparently in cases where the anomaly is very slight and difficult to be identified.


The anomaly detection strategy using the adjacency matrix $A$ presents worse results as the mixture parameter $\mu$ increases. In contrast, the performance improves as the value of $\mu$ increases when \textit{SpecF} employs matrix $W$. This behavior can be explained by the characteristics of the anomalous signal studied in this paper. We analyzed a signal whose anomaly is defined as a vertex whose signal value is different from that of the others in the same community. If the percentage of inter-community edges is low, the average signal values between communities are more distant. Therefore, even if the signal at a node differs from the community where it is located, it will still be closer to the average value of that community than to any other community's average value. When the mixture parameter is higher, the average values of the communities are closer. Moreover, a node whose signal value differs from the signal of its community may be close to the average values of neighboring communities. This characteristic is reflected in the Fourier transform through the expanded adjacency matrix and, then, demonstrated in the applied filter.

\mariah{The results for experiments I, II and III are clearly superior for all compared scenarios in favor of the usage of an expanded adjacency matrix. Therefore, statistical tests like Wilcoxon signed-rank test \cite{demvsar2006statistical} attest the superior performance of the expanded adjacency matrix when compared to the standard counterpart, at 99\% of significance level. }

\subsection{Experiment IV}

\mariah{This experiment employs two publicly available databases. The first, SWaT, contains a water treatment testbed  collected by a sensor-actuator network with 51 nodes \citep{mathur2016swat}. The SWaT dataset regards data collected of a total of 15 days of the water distribution operation, being 11 of normal operations and 4 of day operations  with cyber-attacks. This dataset has an anomaly rate of 5.77\%. The second, called WADI, contains a water distribution testbed  collected by a sensor-actuator network with 123 nodes \citep{ahmed2017wadi}. The duration of the attacks was from 2 to 25 minutes. The WADI dataset regards data collected from a total of 16 days of the water distribution operation, being 14 of normal operation days and 2 with cyber-attacks. The duration of the attacks was from 1.5 to 30 minutes. This dataset has an anomaly rate of 5.75\%.

A comparative analysis is carried out by presenting the results reported in \cite{chen2021learning}. The authors tested their deep learning-based algorithm, GTA, designed to detect anomalies in multivariate time series data. GTA defines anomalies according to a novel graph learning strategy proposed by the authors referred to as Influence Propagation convolution. They compared GTA with eight state-of-the-art anomaly algorithms in multivariate time series data: Principal Component Analysis (PCA), $k$-Nearest Neighbor (KNN) \citep{angiulli2002fast}, Feature Bagging (FB) \citep{lazarevic2005feature},  Autoencoders (AE) \citep{Aggarwal2013}, Long Short Term Memory-based Variational Autoencoder (LSTM-VAE) \citep{park2018multimodal}, Multivariate Anomaly Detection  with Generative Adversarial Networks (MAD-GAN) \citep{li2019mad}, Deep Autoencoding Gaussian Mixture Model (DAGMM) \citep{zong2018deep} and Graph Deviation Network (GDN) \citep{deng2021graph}. 

WADI and SWaT were modeled in two graphs with 112 and 51 vertices, respectively, where the vertices correspond to the sensor nodes and the edges between them contain the Pearson correlation between the values measured by the sensors over time. An edge only exists if the pairwise correlation is greater than 0.5.  

As \textit{SpecF} is not specifically designed to approach time series data, a methodology to define the anomalous signals according to an analysis of pairwise consecutive 30-second time windows of the time series is followed as explained next.

\begin{itemize}
    \item Given a pair of consecutive time windows, calculate their distance using the Dynamic Time Warping (DTW) \cite{sakoe1971dynamic} with the implementation proposed by \citep{wannes_meert_2020_3981067}.
    \item For the two time windows, consider the DTW the signal on the node and apply \textit{SpecF}.
    \item The anomalous nodes  returned by \textit{SpecF} are deemed anomalous in the entire analyzed time windows.
\end{itemize}

Tables \ref{tab:experiment3.2.1} and \ref{tab:experiment3.2.2} present the Precision, Recall and F1-Score of the results obtained by these algorithms and \textit{SpecF} on SWaT and WADI datasets, respectively. The F1-score measure is given Equation~\eqref{f1score}.

\begin{equation}
    \mbox{F1-score}=2\times \frac{\mbox{Precision}\times \mbox{Recall}}{\mbox{Precison+Recall}} \label{f1score}
\end{equation}

\begin{table}[t]
\begin{center}
\caption{Results of anomaly detection algorithms on SWaT dataset.}
\label{tab:experiment3.2.1}
\begin{tabular}{@{} *4l @{}}   \toprule
\emph{Methods} & Precision (\%)  & Recall(\%)  & F1-score  \\\midrule
PCA & 24.92 & 21.63 & 0.23  \\
KNN & 7.83 & 7.83 & 0.08  \\
FB & 10.17 & 10.17 & 0.10  \\
AE & 72.63 & 52.63 & 0.61  \\
DAGMM & 27.46 & 69.52 & 0.39  \\
LSTM-VAE & 96.24 & 59.91 & 0.74  \\
MAD-GAN & 98.97 & 63.74 & 0.77  \\
GDN & 99.35 & 68.12 & 0.81  \\
GTA & 74.91 & 96.41 & 0.84  \\
\hline 
\textit{SpecF} & 55.54 & 62.96 & 0.47  \\
\hline
\end{tabular}

\end{center}
\end{table}

\begin{table}[t]
\begin{center}
\caption{Results of anomaly detection algorithms on WADI dataset.}
\label{tab:experiment3.2.2}
\begin{tabular}{@{} *4l @{}}   \toprule
\emph{Methods} & Precision (\%)  & Recall(\%)  & F1-score  \\\midrule
PCA & 39.53 & 5.63 & 0.10  \\
KNN & 7.76 & 7.75 & 0.08  \\
FB & 8.60 & 8.60 & 0.09  \\
AE & 34.35 & 34.35 & 0.34  \\
DAGMM & 54.44 & 26.99 & 0.36  \\
LSTM-VAE & 87.79 & 14.45 & 0.25  \\
MAD-GAN & 41.44 & 33.92 & 0.37  \\
GDN & 97.50 & 40.19 & 0.57  \\
GTA & 74.56 & 90.50 & 0.82  \\
\hline 
\textit{SpecF} & 49.88 & 49.44 & 0.35  \\
\hline
\end{tabular}

\end{center}
\end{table}

According to the F1-score, \textit{SpecF} outperformed the PCA, KNN, FB and DAGMM methods in both datasets. In the WADI dataset, \textit{SpecF} also outperformed the LSTM-VAE method.  In contrast, \textit{SpecF} was outperformed by MAD-GAN, GDN and GTA. Nonetheless, different from these methods, \textit{SpecF} does not need any training steps to obtain information on the non-anomalous state of the system. Therefore, we believe that \textit{SpecF} presented a good performance and was able to produce satisfactory results with a low computational cost. Although it was not specifically designed to deal with time series, the results point out that \textit{SpecF} achieved good results, being a good option to identify anomalies in multiple time series that show some correlation.}

\subsection{Case Study - COVID-19 dataset}
\begin{figure}
	\centering
		\includegraphics[scale=0.12]{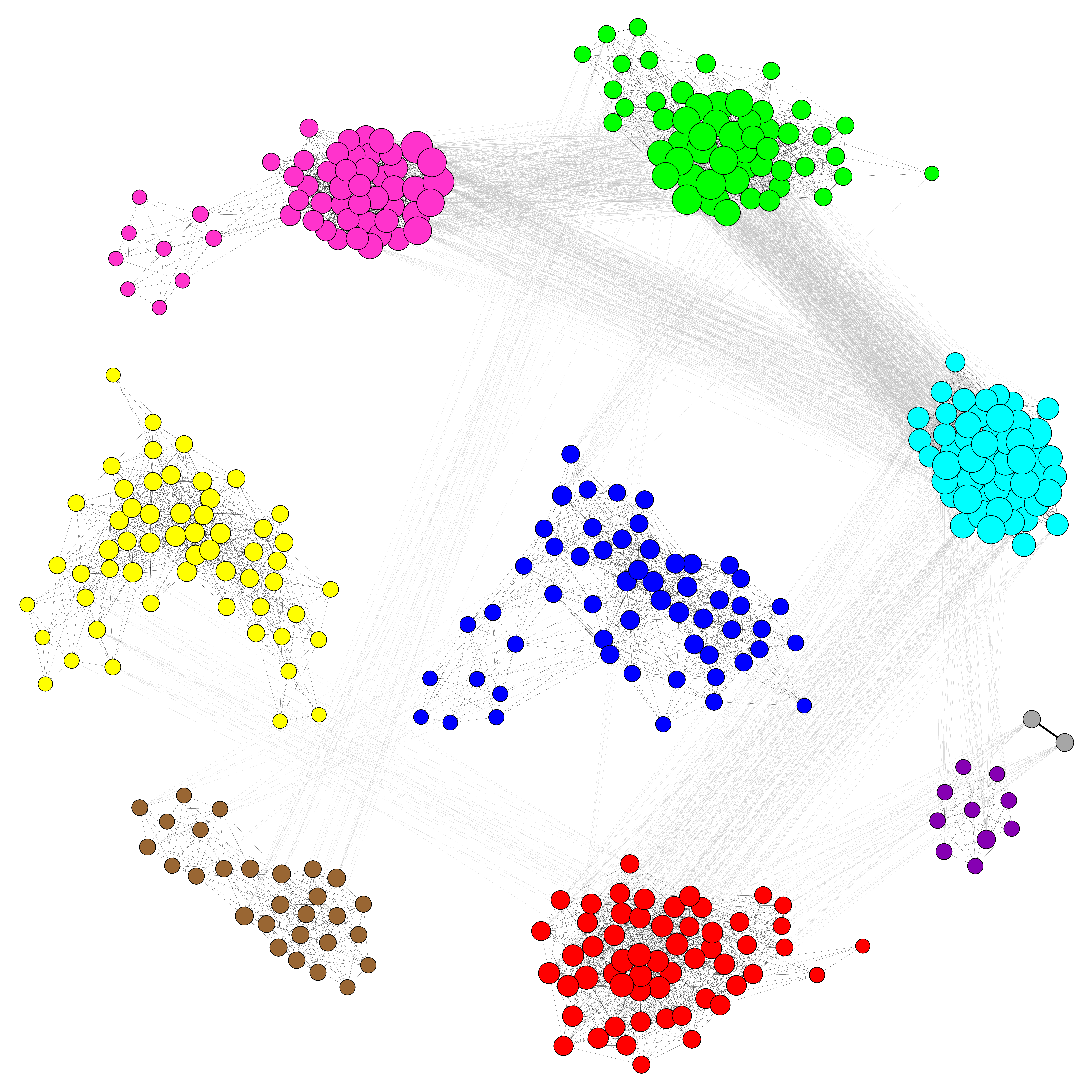}
	\caption{Graph representing the districts of São José dos Campos, SP, Brazil.}
	\label{fig:graph}
\end{figure}

Here an experiment with real data to validate the performance of \textit{SpecF} was carried out. The data set used in the experiment contains the number of daily  COVID-19  cases in each of the districts of São José dos Campos, an upstate city of São Paulo - Brazil. To model the data, let $G$ be a graph where the nodes represent districts of the city. A pair of nodes $v_i$ and $v_j$ is connected if the distance between the corresponding districts is less than 5 km. In this case, the weight $w_{ij}$ of the edge is proportional to the distance between districts represented by nodes $v_i$ and $v_j$. Figure \ref{fig:graph} illustrates the graph $G$, where the nodes were colored according to their communities (meaning that nodes with identical color belong to the same community). The case study graph has 366 nodes and the community structure of the network was identified using a community detection algorithm named Louvain \citep{Blondel2008FastNetworks}. The Louvain method is a benchmark community detection heuristic that identifies the community structure through the modularity optimization  \citep{Newman2004FindingNetworks}.

The data used was collected from March 5, 2020 to March 25, 2021, totaling 385 days (equivalent to 55 weeks). For each week, only the nodes that represented districts that presented at least 2 cases in the previous and current weeks are included. Thus, for each week, the number of nodes and the community structure of the graph are modified due to the removal of some nodes that represent districts that did not have a sufficient number of cases in the analyzed weeks. We define the signal $B^{(d)}$ as the weekly variation in the number of COVID-19 cases, where $b_i^{(d)}$ is the ratio between the sum of the total number of cases registered between days $d-7, d-6, \ldots, d-1$ and days $d, d+1, \ldots, d+6$ in the district represented by node $v_i$. The proposed anomaly detection strategy was applied to signal $B^{(d)}$ to identify anomalies in the weekly variation of cases. The hypothesis is that if the number of cases in a given district increases, the number of cases in adjacent districts must increase as well. As a consequence, \textit{SpecF} may identify neighbor districts whose variation in the number of cases differs from a given district in a certain week.

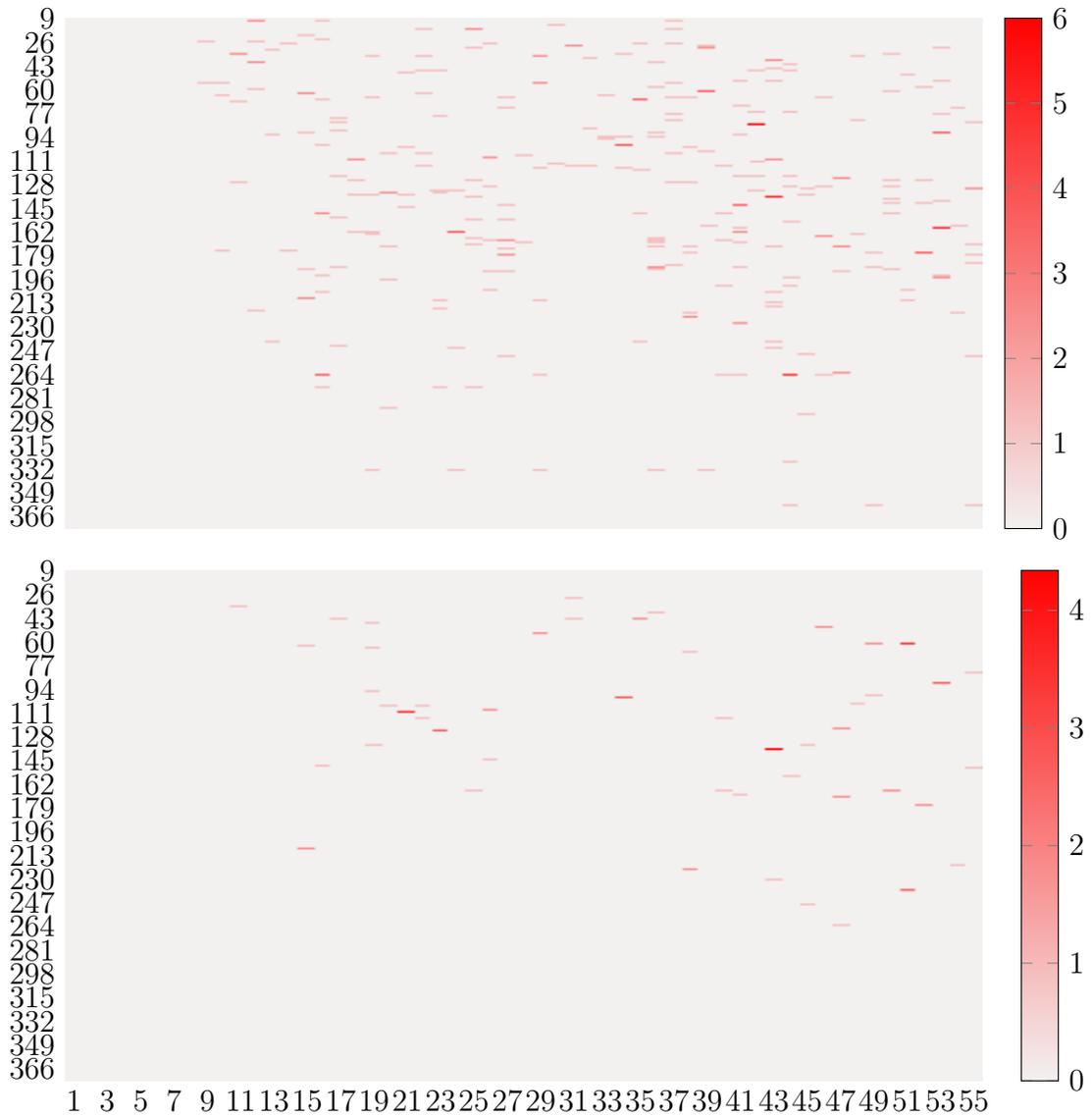
\begin{figure}[h!t]
\center

\input{tikz/hitmap.tex}

\caption{Weekly variation in the number of cases in each of the districts of São José dos Campos, SP, Brazil.}
	\label{fig:hitmap}
\end{figure}

The upper part of Figure \ref{fig:hitmap} illustrates the weekly variation in the number of cases in each of the districts. The 366 different districts are represented by the y-axis whereas the x-axis represents the weeks. The signal corresponding to week 2, for example, is represented by $B^{(14)}$, since $d$ is reached by multiplying the week number by 7. The intensity of the colors represents the weekly variation in the number of cases. The redder the greater the increase in the weekly number of cases. The modeled graph is subject to \textit{SpecF} to find the $Y^{(d)}$ signal that stores the abnormality of the nodes. The lower part of Figure \ref{fig:hitmap} illustrates the abnormality in the variation on the number of cases for each district. It is possible to notice that the amount of light dots at the bottom is considerably smaller. This is because \textit{SpecF} filters the signal and highlights the abnormalities, so that only potentially anomalous variations are maintained in the $Y^{(d)}$ signal.


To validate the hypothesis behind the introduced anomaly detection algorithm and attest to the veracity of the result, we compared the values of the nodes identified as anomalous with the values of the other nodes of the same community for every day $d = \{0,7,14, \ldots,378, 385\}$. For example, if $y_i^{(d)}$ is pointed out as an anomaly, it means that the variation in the number of cases between the days $d-7,d-1$ and $d,d+6$ in the district represented by node $v_i$ differs from the standard of the community to which $v_i$ node belongs. 

\begin{figure}
	\centering
		\includegraphics[scale=.24]{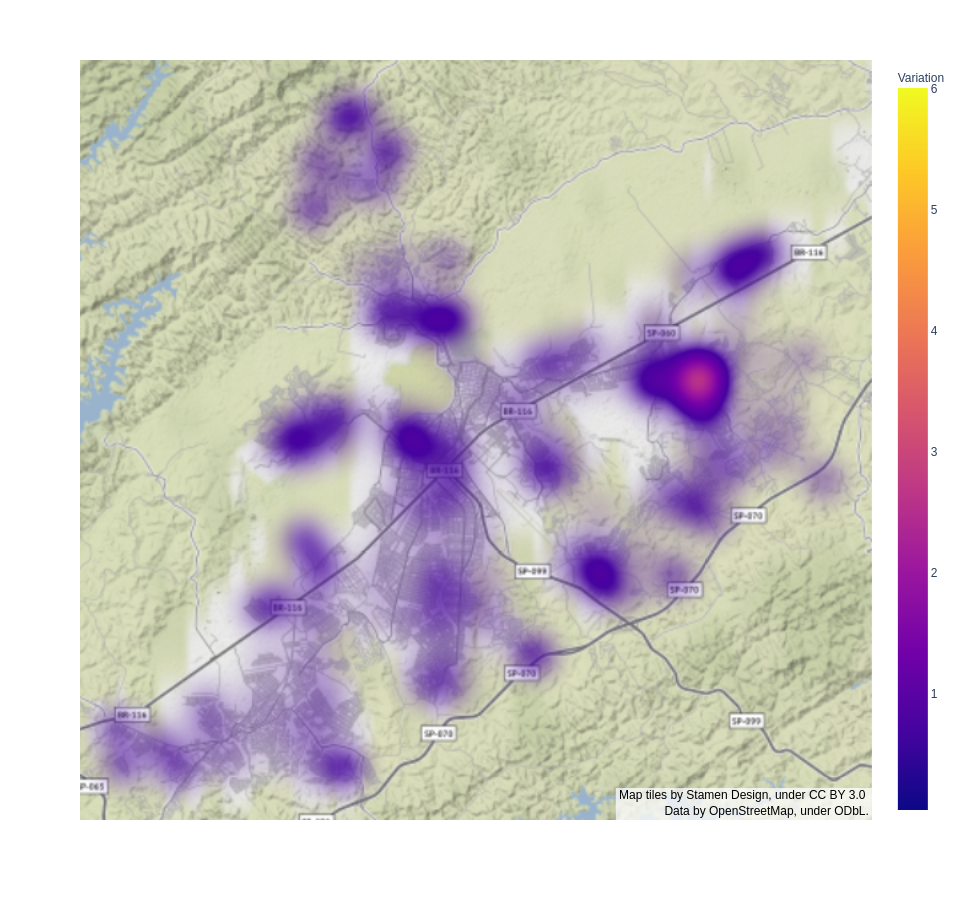}
	\caption{Heat map with the weekly variation in the number of cases in each of the district of São José dos Campos, SP, Brazil.}
	\label{fig:map}
\end{figure}

This behavior is best seen in Figure \ref{fig:map}, which illustrates a map with the weekly variation in the number of cases within the districts, with a single anomaly. Shades closer to blue represent a drop in the number of cases, while shades closer to yellow represent an increase in the total of COVID-19 cases. It is possible to observe that most districts experience a reduction in the number of cases, while only one district has an increase. This increase is considered an abnormal incident since the adjacent districts have the number of COVID-19 cases reduced. 

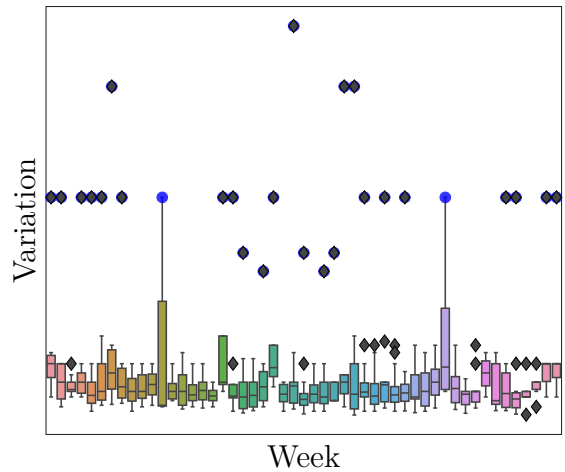
\begin{figure}[h!t]
\center

\input{tikz/boxplot.tex}

\caption{Box-plots of the weekly variation in the number of COVID-19 cases considering communities with anomalous nodes.}
	\label{fig:boxplot}
\end{figure}

A total of 117 values $y_i^{(d)} $ were identified as anomalous. Approximately 95\% of these anomalies have the same characteristic: the node identified as anomalous in a time window was the only one that had an increase in the number of cases in its community. \mariah{This highlights neighborhoods and routes of dissemination of the virus that may require public health interventions.}
 Figure \ref{fig:boxplot} shows this behavior for a sample of 50 of the 117 anomalies identified. Each box-plot illustrates the variation values of nodes from the same community to which the anomalous node belongs. The variation values of the anomalous nodes are highlighted in blue dots. The anomalous node presents an increase in the number of COVID-19 cases, in contrast to the other nodes of the same community, that showed a reduction in the number of cases. 

We have identified that new anomalies usually occur in districts that are adjacent to districts where abnormal variations were observed in the previous two weeks. In 80.3\% of the cases, if a node $v_i$ is indicated as anomalous in the signal $y_i^{(d)}$, then there is a node $v_j$, neighbor of $v_i$, that is pointed out as anomalous in the signal $y_j^{(d-7)}$ or $y_j^{(d-14)}$, where $d\geq 14$. The results obtained in this case study indicate that our hypothesis was correct. The proposed strategy can identify nodes whose characteristics differ from the others from the same community, even in real-world scenarios. Furthermore, it shows that the anomaly trails through the network, such that an anomalous node hits adjacent nodes that become anomalous a few weeks later. \mariah{Such analyzes can support anticipated public health decisions for a better management of the pandemics evolution in the city.}

\section{Conclusions}\label{sec:c}

This paper proposed a method to detect anomalies in signals indexed by graphs, called \textit{SpecF}. \textit{SpecF} uses an extended Laplacian matrix that considers the community structure as the basis for the Fourier transform and a low-pass filter designed to attenuate high frequencies in the signal spectrum. The proposed filter uses as cut-off frequency the eigenvalue whose position is the number of communities in the network. This position can be passed as a parameter or automatically calculated through eigenvalue analysis of the adopted matrix.

Computational experiments attest to the accuracy of \textit{SpecF} in the task of detecting anomalies. Moreover, the automatic calculation of the cut-off frequency using the network community structure makes the method independent of parameters and more robust when dealing with real-world networks. In addition, the experiments show that the proposed method performed better when adopting the expanded adjacency matrix $W$ in the Fourier transform instead of the standard adjacency matrix $A$.

The results suggest that \textit{SpecF} is an interesting solution to detect anomalies based on signals indexed by graphs. More specifically, the method can successfully identify nodes in the networks whose signal values differ from the expected according to the community where they belong. Therefore, the proposed method effectively finds anomalies in signals whose expected variation is low in nodes from the same community. \mariah{Moreover, a comparative analysis  with state-of-the-art algorithms in two publicly available real world datasets with known anomalies show a good performance of \textit{SpecF}. \textit{SpecF} outperformed the classic algorithms, being competitive with some recent reference methods, even though it has not any learning step and has not been designed for the time series anomaly detection task. }

Additionally, a case study was presented to validate the proposed strategy. It consists of data related to the number of COVID-19 cases in the different districts of São José dos Campos, an upstate city of São Paulo, Brazil. By using \textit{SpecF}, it was possible to detect districts whose number of cases soared while neighboring districts showed a reduction in the number of cases in the same week. The number of COVID-19 cases in a district in a given week was evaluated abnormally high or low. The number of COVID-19 cases in some neighbor districts in the former one or two weeks is also anomalous in most of the cases. This gives evidence that the uncontrolled dissemination of the virus occurs following a ``path'' in the city. 

\section{Acknowledgments}

The authors of this paper are grateful to Fundação de Amparo à Pesquisa do Estado de São Paulo (FAPESP) (Grant Number: 2017/24185-0), Coordenação de Aperfeiçoamento de Pessoal de Nível Superior - Brasil (CAPES)  (Grant: 88887.507037/2020-00) and to the health department of São José dos Campos for providing the COVID-19 data.

\bibliographystyle{elsarticle-harv} 
\bibliography{references}

\end{document}

%% file: packages.tex
\usepackage[T1]{fontenc}

\usepackage[utf8]{inputenc}

\usepackage{graphicx}
\usepackage{epstopdf}
\usepackage{amssymb}
\usepackage{amsmath,amsthm}
\usepackage{mathtools}
\usepackage{commath}
\usepackage{booktabs}
\usepackage[ruled,vlined]{algorithm2e}
\usepackage{caption}
\usepackage{svg}
\usepackage{tabularx}
\usepackage{lmodern,amsfonts,mathrsfs,amsmath,amssymb,latexsym,xspace,epsfig,psfrag,color,tikz,tikzscale,pgfplots,colortbl}
 \usepackage{amsmath}

\usepackage{tikzscale}
\usepackage{tikz,pgfplots}
\usepgfplotslibrary{statistics}

\usepackage[utf8]{inputenc}
\usepackage{pgfplots}
\DeclareUnicodeCharacter{2212}{−}
\usepgfplotslibrary{groupplots,dateplot}
\usetikzlibrary{patterns,shapes.arrows}
\pgfplotsset{compat=newest}

\newcommand{\mathbbm}[1]{\text{\usefont{U}{bbm}{m}{n}#1}} 

\newcommand\MyBox[2]{
  \fbox{\lower0.45cm
    \vbox to 1.0cm{\vfil
      \hbox to 0.9cm{\hfil\parbox{1.0cm}{#1}\hfil}
      \vfil}
  }
}

\newcommand\ddfrac[2]{\frac{\displaystyle #1}{\displaystyle #2}}

\journal{Applied Soft Computing}

%% file: tikz/anomalous_ordered2.tex
\begin{tikzpicture}

\begin{axis}[
legend cell align={left},
legend style={
  fill opacity=0.8,
  draw opacity=1,
  text opacity=1,
  at={(0.03,0.97)},
  anchor=north west,
  draw=white!80!black
},
xlabel={\(\displaystyle V\)},
xmin=-5, xmax=505,
xtick={-100,0,100,200,300,400,500,600},
xticklabel style={rotate=90.0},
grid=major,
ylabel={\(\displaystyle B'\)},
ymin=12123.3591995191, ymax=678055.744213591,
]
\addplot [
  draw=white,
  mark=*,
  color=blue,
  only marks,
  mark size=1pt
]
table[x=x,y=y] {tikz/anomalous_ordered2_1.txt};
\addlegendentry{Normal}
\addplot [
  draw=white,
  mark=*,
  red,
  only marks,
  mark size=1pt
]
table[x=x,y=y] {tikz/anomalous_ordered2_2.txt};
\addlegendentry{Anomaly}
\end{axis}

\end{tikzpicture}

%% file: tikz/anomalous_type.tex
\begin{tikzpicture}

\begin{axis}[
legend cell align={left},
legend style={
  fill opacity=0.8,
  draw opacity=1,
  text opacity=1,
  at={(0.03,0.97)},
  anchor=north west,
  draw=white!80!black
},
xlabel={\(\displaystyle V\)},
xmin=-5, xmax=505,
xtick={-100,0,100,200,300,400,500,600},
xticklabel style={rotate=90.0},
grid=major,
ylabel={\(\displaystyle Y\)},
ymin=0, ymax=150000,
]
\addplot [
  draw=white,
  mark=*,
  color=blue,
  only marks,
  mark size=1pt
]
table[x=x,y=y] {tikz/anomalous_type_1.txt};
\addlegendentry{true negative}
\addplot [
  draw=white,
  mark=*,
  red,
  only marks,
  mark size=1pt
]
table[x=x,y=y] {tikz/anomalous_type_2.txt};
\addlegendentry{true positive}
\addplot [
  draw=white,
  mark=*,
  green,
  only marks,
  mark size=1pt
]
table[x=x,y=y] {tikz/anomalous_type_3.txt};
\addlegendentry{false positive}
\addplot [
  draw=white,
  mark=*,
  yellow,
  only marks,
  mark size=1pt
]
table[x=x,y=y] {tikz/anomalous_type_4.txt};
\addlegendentry{false negative}

\addplot [
  draw=white,
  mark=*,
  dashed,
  gray,
  mark size=0.1pt
]
table[x=x,y=y] {tikz/anomalous_type_5.txt};
\addlegendentry{threshold}
\end{axis}

\end{tikzpicture}

%% file: tikz/box_GLOBAL.tex
\definecolor{RYB1}{RGB}{230,97,1}
\definecolor{RYB2}{RGB}{200,150,250}
\pgfplotscreateplotcyclelist{colorbrewer-RYB}{
{blue!60!white,fill=blue!30!white,line width=1.5pt},
{red!70!white,fill=red!30!white,line width=1.5pt},
{orange!70!white,fill=orange!30!white,line width=1.5pt},
{violet!70!white,fill=violet!30!white,line width=1.5pt}}
\begin{tikzpicture}
\pgfplotsset{every axis legend/.append style={
at={(0.8,0.05)},
anchor=south}
},
\begin{axis}[
width=12cm,
height=7.5cm,
ymin=0,
boxplot/draw direction=y,
cycle list name=colorbrewer-RYB,
enlarge y limits,
ymajorgrids,
xlabel={\(\displaystyle Community\)},
ylabel={\(\displaystyle S\)},
xtick={1,2,3,4,5,6,7,8,9,10},
xtick style={draw=none},
xticklabels={$C_1$,$C_4$,$C_3$,$C_2$,$C_6$,$C_5$,$C_9$,$C_8$,$C_7$,$C_{10}$},
 legend style={at={(0.5,1.15)}, anchor=north,legend columns=-1},
    legend image code/.code={
    \draw[#1, draw=none] (0cm,-0.1cm) rectangle (0.6cm,0.1cm);
    },
]
\addplot+[
boxplot prepared={
draw position=1,
lower whisker=12761.430736335884, lower quartile=34813.03590860689,
median=40761.33683575776, 
upper quartile=54839.75477360727, upper whisker=89163.73577490078,
},
]
coordinates {};

\addplot+[
boxplot prepared={
draw position=2,
lower whisker=47930.408866102924, lower quartile=59536.53971021194,
median=72973.26059373398, 
upper quartile=82123.94681215685, upper whisker=135681.65087689884,
},
]
coordinates {};

\addplot+[
boxplot prepared={
draw position=3,
lower whisker=136851.3836720383, lower quartile=145502.3589184699,
median=150594.37386486452, 
upper quartile=163401.96519980818, upper whisker=186646.45491574,
},
]
coordinates {};

\addplot+[
boxplot prepared={
draw position=4,
lower whisker=223069.959801411, lower quartile=243130.97635084932,
median=259820.61725857545, 
upper quartile=273866.94260095275, upper whisker=326551.6848796279,
},
]
coordinates {};

\addplot+[
boxplot prepared={
draw position=5,
lower whisker=288141.5848444893, lower quartile=313985.3236074458,
median=327035.5573585767, 
upper quartile=340143.58515518147, upper whisker=421994.0798393728,
},
]
coordinates {};

\addplot+[
boxplot prepared={
draw position=6,
lower whisker=348966.9317520538, lower quartile=376752.7915247908,
median=393034.8473749313, 
upper quartile=410986.2088748581, upper whisker=479054.5512365435,
},
]
coordinates {};

\addplot+[
boxplot prepared={
draw position=7,
lower whisker=347196.91258856206, lower quartile=379451.8096820041,
median=396560.77431748615, 
upper quartile=410339.4085187771, upper whisker=478356.37824863236,
},
]
coordinates {};

\addplot+[
boxplot prepared={
draw position=8,
lower whisker=411550.27564390114, lower quartile=441892.01911851036,
median=460747.2765698683, 
upper quartile=477935.2130007492, upper whisker=547051.1253478745,
},
]
coordinates {};

\addplot+[
boxplot prepared={
draw position=9,
lower whisker=424436.6154330159, lower quartile=483215.1595523226,
median=508797.6827631482, 
upper quartile=544137.1644101182, upper whisker=645767.3754415149,
},
]
coordinates {};

\addplot+[
boxplot prepared={
draw position=10,
lower whisker=470220.3468409204, lower quartile=502401.07063466817,
median=523310.547945073, 
upper quartile=555481.1428911859, upper whisker=658329.6166409694,
},
]
coordinates {};

\end{axis}
\end{tikzpicture}

%% file: tikz/anomalous_ordered.tex
\begin{tikzpicture}

\begin{axis}[
legend cell align={left},
legend style={
  fill opacity=0.8,
  draw opacity=1,
  text opacity=1,
  at={(0.03,0.97)},
  anchor=north west,
  draw=white!80!black
},
xlabel={\(\displaystyle V\)},
xmin=-5, xmax=505,
xtick={-100,0,100,200,300,400,500,600},
xticklabel style={rotate=90.0},
grid=major,
ylabel={\(\displaystyle B\)},
ymin=12123.3591995191, ymax=678055.744213591,
]
\addplot [
  draw=white,
  mark=*,
  color=blue,
  only marks,
  mark size=1pt
]
table[x=x,y=y] {tikz/anomalous_ordered_1.txt};
\addlegendentry{Normal}
\addplot [
  draw=white,
  mark=*,
  red,
  only marks,
  mark size=1pt
]
table[x=x,y=y] {tikz/anomalous_ordered_2.txt};
\addlegendentry{Anomaly}
\end{axis}

\end{tikzpicture}

%% file: tikz/output.tex
\begin{tikzpicture}

\definecolor{color0}{rgb}{0.12156862745098,0.466666666666667,0.705882352941177}
\definecolor{color1}{rgb}{1,0.498039215686275,0.0549019607843137}

\begin{axis}[
axis line style={white!15!black},
legend cell align={left},
legend style={fill opacity=0.8, draw opacity=1, text opacity=1, draw=white!80!black, at={(1.5,1)},anchor=north},
xlabel={\(\displaystyle \mu\)},
xmin=-0.35, xmax=7.35,
xtick={0,1,2,3,4,5,6,7},
xticklabels={0.1,0.2,0.3,0.4,0.5,0.6,0.7,0.8},
ylabel={Mean AP},
ymin=0.133235170932329, ymax=0.952236577735355,
ytick={0.1,0.2,0.3,0.4,0.5,0.6,0.7,0.8,0.9,1},
yticklabels={0.1,0.2,0.3,0.4,0.5,0.6,0.7,0.8,0.9,1.0}
]
\path [draw=color0, fill=color0, opacity=0.2]
(axis cs:0,0.527384631410822)
--(axis cs:0,0.390693271903608)
--(axis cs:1,0.326508437667071)
--(axis cs:2,0.306137015132359)
--(axis cs:3,0.25088079682969)
--(axis cs:4,0.226360514273647)
--(axis cs:5,0.184812109677948)
--(axis cs:6,0.170462507605194)
--(axis cs:7,0.195865504326926)
--(axis cs:7,0.308874976273907)
--(axis cs:7,0.308874976273907)
--(axis cs:6,0.274724097446817)
--(axis cs:5,0.280738185788255)
--(axis cs:4,0.341450216867912)
--(axis cs:3,0.378348036675038)
--(axis cs:2,0.451046784088869)
--(axis cs:1,0.474156091736106)
--(axis cs:0,0.527384631410822)
--cycle;

\path [draw=color0, fill=color0, opacity=0.2]
(axis cs:0,0.586376665936274)
--(axis cs:0,0.467418038540205)
--(axis cs:1,0.385223929757714)
--(axis cs:2,0.390007445162739)
--(axis cs:3,0.310076300643118)
--(axis cs:4,0.287411995129978)
--(axis cs:5,0.21226732400618)
--(axis cs:6,0.23054253434272)
--(axis cs:7,0.272530262914397)
--(axis cs:7,0.384595703698388)
--(axis cs:7,0.384595703698388)
--(axis cs:6,0.336482295198494)
--(axis cs:5,0.333893197146265)
--(axis cs:4,0.403825028780971)
--(axis cs:3,0.441015454610602)
--(axis cs:2,0.519509388588852)
--(axis cs:1,0.527224886590866)
--(axis cs:0,0.586376665936274)
--cycle;

\path [draw=color0, fill=color0, opacity=0.2]
(axis cs:0,0.6457345438823)
--(axis cs:0,0.537288803426067)
--(axis cs:1,0.444200119712263)
--(axis cs:2,0.473812178665318)
--(axis cs:3,0.401377137036693)
--(axis cs:4,0.351391810748481)
--(axis cs:5,0.30092008658311)
--(axis cs:6,0.30362573078536)
--(axis cs:7,0.371226222678587)
--(axis cs:7,0.506730473824603)
--(axis cs:7,0.506730473824603)
--(axis cs:6,0.422824925239015)
--(axis cs:5,0.409136704338963)
--(axis cs:4,0.478753037230774)
--(axis cs:3,0.520672381087529)
--(axis cs:2,0.587697451487266)
--(axis cs:1,0.584983023406634)
--(axis cs:0,0.6457345438823)
--cycle;

\path [draw=color1, fill=color1, opacity=0.2]
(axis cs:0,0.565157104696492)
--(axis cs:0,0.404571485027695)
--(axis cs:1,0.418980006245827)
--(axis cs:2,0.49439535895842)
--(axis cs:3,0.500582841256637)
--(axis cs:4,0.531439675305293)
--(axis cs:5,0.527798916755546)
--(axis cs:6,0.464737066978756)
--(axis cs:7,0.396987447202461)
--(axis cs:7,0.56484128631923)
--(axis cs:7,0.56484128631923)
--(axis cs:6,0.622034484050736)
--(axis cs:5,0.663298084053583)
--(axis cs:4,0.671673994663461)
--(axis cs:3,0.648276712691146)
--(axis cs:2,0.649649559008643)
--(axis cs:1,0.589442059295814)
--(axis cs:0,0.565157104696492)
--cycle;

\path [draw=color1, fill=color1, opacity=0.2]
(axis cs:0,0.620316014323981)
--(axis cs:0,0.467013103900298)
--(axis cs:1,0.495913429767379)
--(axis cs:2,0.585909790056815)
--(axis cs:3,0.615355909838702)
--(axis cs:4,0.68283412135801)
--(axis cs:5,0.722839190478696)
--(axis cs:6,0.634656357189933)
--(axis cs:7,0.62011819437288)
--(axis cs:7,0.748565877244541)
--(axis cs:7,0.748565877244541)
--(axis cs:6,0.781378073079429)
--(axis cs:5,0.813148184026965)
--(axis cs:4,0.796940601874984)
--(axis cs:3,0.751166096860871)
--(axis cs:2,0.725980908352376)
--(axis cs:1,0.64837603418753)
--(axis cs:0,0.620316014323981)
--cycle;

\path [draw=color1, fill=color1, opacity=0.2]
(axis cs:0,0.672000917927961)
--(axis cs:0,0.577413761559138)
--(axis cs:1,0.586390209044519)
--(axis cs:2,0.670676225071972)
--(axis cs:3,0.719410729569391)
--(axis cs:4,0.833972386456401)
--(axis cs:5,0.863957651470098)
--(axis cs:6,0.821023057813502)
--(axis cs:7,0.849126636048366)
--(axis cs:7,0.900904975207691)
--(axis cs:7,0.900904975207691)
--(axis cs:6,0.908262661511611)
--(axis cs:5,0.91500924106249)
--(axis cs:4,0.890116642674704)
--(axis cs:3,0.839360218941052)
--(axis cs:2,0.792058869822312)
--(axis cs:1,0.706473373413659)
--(axis cs:0,0.672000917927961)
--cycle;

\addplot [semithick, color0, mark=*, mark size=3, mark options={solid,draw=white}]
table {%
0 0.463889598846436
1 0.405169248580933
2 0.386870861053467
3 0.321290969848633
4 0.286810994148254
5 0.234594583511353
6 0.225273013114929
7 0.250971555709839
};
\addlegendentry{$M_G= A$, $\theta$ = 1\%}

\addplot [semithick, color0, dash pattern=on 4pt off 1.5pt, mark=x, mark size=3, mark options={solid,draw=white}]
table {%
0 0.530823111534119
1 0.463309168815613
2 0.457791328430176
3 0.379536628723145
4 0.347146272659302
5 0.275834798812866
6 0.282111406326294
7 0.330282092094421
};
\addlegendentry{$M_G= A$, $\theta$ = 5\%}

\addplot [semithick, color0, dash pattern=on 1pt off 1pt]
table {%
0 0.595852375030518
1 0.51937460899353
2 0.533390998840332
3 0.467111468315125
4 0.421446800231934
5 0.356431722640991
6 0.367942333221436
7 0.441259741783142
};
\addlegendentry{$M_G= A$, $\theta$ = 10\%}

\addplot [semithick, color1, mark=*, mark size=3, mark options={solid,draw=white}]
table {%
0 0.487389087677002
1 0.508917450904846
2 0.581488370895386
3 0.582687377929688
4 0.609492421150208
5 0.604689836502075
6 0.549417495727539
7 0.488889098167419
};
\addlegendentry{$M_G= W$, $\theta$ = 1\%}

\addplot [semithick, color1, dash pattern=on 4pt off 1.5pt, mark=x, mark size=3, mark options={solid,draw=white}]
table {%
0 0.549430727958679
1 0.580176830291748
2 0.660313248634338
3 0.690369725227356
4 0.745954751968384
5 0.771763324737549
6 0.71807873249054
7 0.689541101455688
};
\addlegendentry{$M_G= W$, $\theta$ = 5\% }

\addplot [semithick, color1, dash pattern=on 1pt off 1pt]
table {%
0 0.630645751953125
1 0.65432071685791
2 0.740049600601196
3 0.78788423538208
4 0.86527419090271
5 0.89130973815918
6 0.871292352676392
7 0.876066565513611
};
\addlegendentry{$M_G= W$, $\theta$ = 10\% }


\end{axis}

\end{tikzpicture}

%% file: tikz/hitmap.tex
\pgfplotsset{width=14cm,height=8.5cm}
\begin{tikzpicture}

\definecolor{color0}{rgb}{0.917647058823529,0.917647058823529,0.949019607843137}

\begin{axis}[
axis background/.style={fill=color0},
axis line style={white},
colorbar,
colorbar style={ylabel={}},
colormap={mymap}{[1pt]
 rgb(0pt)=(0.952263507537876,0.941107372843543,0.940983659029102);
  rgb(1pt)=(0.961810806030301,0.752885898274835,0.752786927223282);
  rgb(2pt)=(0.971358104522726,0.564664423706126,0.564590195417461);
  rgb(3pt)=(0.980905403015151,0.376442949137417,0.376393463611641);
  rgb(4pt)=(0.990452701507575,0.188221474568709,0.18819673180582);
  rgb(5pt)=(1,0,0)
},
point meta max=6,
point meta min=0,
x grid style={white},
xmajorticks=false,
xmin=0, xmax=55,
y dir=reverse,
y grid style={white},
ymin=0, ymax=366,
ytick={0.5,17.5,34.5,51.5,68.5,85.5,102.5,119.5,136.5,153.5,170.5,187.5,204.5,221.5,238.5,255.5,272.5,289.5,306.5,323.5,340.5,357.5},
yticklabels={9,26,43,60,77,94,111,128,145,162,179,196,213,230,247,264,281,298,315,332,349,366}
]
\addplot graphics [includegraphics cmd=\pgfimage,xmin=0, xmax=55, ymin=366, ymax=0] {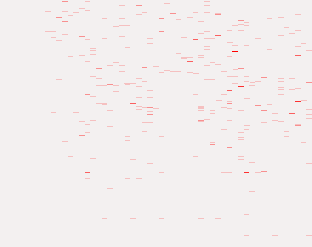};
\end{axis}
\end{tikzpicture}

\begin{tikzpicture}

\definecolor{color0}{rgb}{0.917647058823529,0.917647058823529,0.949019607843137}

\begin{axis}[
axis background/.style={fill=color0},
axis line style={white},
colorbar,
colorbar style={ylabel={}},
colormap={mymap}{[1pt]
 rgb(0pt)=(0.952263507537876,0.941107372843543,0.940983659029102);
  rgb(1pt)=(0.961810806030301,0.752885898274835,0.752786927223282);
  rgb(2pt)=(0.971358104522726,0.564664423706126,0.564590195417461);
  rgb(3pt)=(0.980905403015151,0.376442949137417,0.376393463611641);
  rgb(4pt)=(0.990452701507575,0.188221474568709,0.18819673180582);
  rgb(5pt)=(1,0,0)
},
point meta max=4.33996031589818,
point meta min=0,
x grid style={white},
xmin=0, xmax=55,
xtick={0.5,2.5,4.5,6.5,8.5,10.5,12.5,14.5,16.5,18.5,20.5,22.5,24.5,26.5,28.5,30.5,32.5,34.5,36.5,38.5,40.5,42.5,44.5,46.5,48.5,50.5,52.5,54.5},
xticklabels={1,3,5,7,9,11,13,15,17,19,21,23,25,27,29,31,33,35,37,39,41,43,45,47,49,51,53,55},
y dir=reverse,
y grid style={white},
ymin=0, ymax=366,
ytick={0.5,17.5,34.5,51.5,68.5,85.5,102.5,119.5,136.5,153.5,170.5,187.5,204.5,221.5,238.5,255.5,272.5,289.5,306.5,323.5,340.5,357.5},
yticklabels={9,26,43,60,77,94,111,128,145,162,179,196,213,230,247,264,281,298,315,332,349,366}
]
\addplot graphics [includegraphics cmd=\pgfimage,xmin=0, xmax=55, ymin=366, ymax=0] {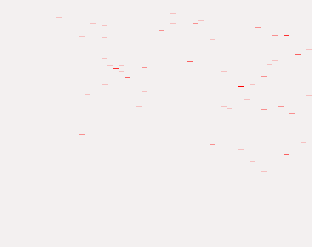};
\end{axis}

\end{tikzpicture}

%% file: tikz/boxplot.tex
\begin{tikzpicture}

\definecolor{color0}{rgb}{0.916118471592631,0.587423760636183,0.640263770354658}
\definecolor{color1}{rgb}{0.917363216417896,0.592149496852113,0.602324495610027}
\definecolor{color2}{rgb}{0.911913257573137,0.590318544894614,0.549797594355991}
\definecolor{color3}{rgb}{0.902392630761742,0.584683156013944,0.478577227016643}
\definecolor{color4}{rgb}{0.887241747660813,0.573849885605829,0.367356737781836}
\definecolor{color5}{rgb}{0.847343840914552,0.578821125309127,0.305309083354541}
\definecolor{color6}{rgb}{0.806908109112525,0.59114836197454,0.299181002756321}
\definecolor{color7}{rgb}{0.771438902812091,0.600306844790949,0.29382413700404}
\definecolor{color8}{rgb}{0.739018079280079,0.607489795702101,0.288942624389942}
\definecolor{color9}{rgb}{0.708261520156588,0.613360866791132,0.284324738049742}
\definecolor{color10}{rgb}{0.678040532298002,0.618314233779748,0.279799476474175}
\definecolor{color11}{rgb}{0.64729674693634,0.622590749098047,0.275208186802333}
\definecolor{color12}{rgb}{0.616786424493649,0.628239939634251,0.272288796615818}
\definecolor{color13}{rgb}{0.587706654203615,0.637970491292984,0.273478183457922}
\definecolor{color14}{rgb}{0.55418089085994,0.647969978771459,0.274695786479355}
\definecolor{color15}{rgb}{0.513262469650651,0.658578617018414,0.275982352646808}
\definecolor{color16}{rgb}{0.459106866331762,0.670221592595001,0.277388118246666}
\definecolor{color17}{rgb}{0.376222260682492,0.6834863779637,0.278981649858852}
\definecolor{color18}{rgb}{0.281435025932605,0.69132943576454,0.343859636342258}
\definecolor{color19}{rgb}{0.284200010413036,0.687197113761566,0.439163662548389}
\definecolor{color20}{rgb}{0.286338761686701,0.683885071734601,0.49490354179113}
\definecolor{color21}{rgb}{0.288089812829126,0.681095048341337,0.534420411690126}
\definecolor{color22}{rgb}{0.289591263449553,0.678644501979596,0.565278955954163}
\definecolor{color23}{rgb}{0.290930533650636,0.676411864776264,0.590986974438807}
\definecolor{color24}{rgb}{0.292167847288054,0.674308866355028,0.613492821994873}
\definecolor{color25}{rgb}{0.293348608536971,0.67226489930959,0.634026483578495}
\definecolor{color26}{rgb}{0.294510751808618,0.670216870219712,0.65345878864305}
\definecolor{color27}{rgb}{0.296420480090951,0.668831788453843,0.673215469101712}
\definecolor{color28}{rgb}{0.301239023054702,0.670160886126314,0.696053008299471}
\definecolor{color29}{rgb}{0.306343814424822,0.671445089587235,0.719965599854794}
\definecolor{color30}{rgb}{0.311946937012235,0.672709520314164,0.745895225872228}
\definecolor{color31}{rgb}{0.318342710097883,0.67396891790319,0.77510866878648}
\definecolor{color32}{rgb}{0.325981404408343,0.675218139707178,0.80949465151157}
\definecolor{color33}{rgb}{0.335621881004181,0.676400626470144,0.852163311844587}
\definecolor{color34}{rgb}{0.426613526825655,0.681400532693827,0.887856226904857}
\definecolor{color35}{rgb}{0.527912637948111,0.6823644022552,0.902256389106443}
\definecolor{color36}{rgb}{0.600419611939353,0.67959921676241,0.912547726004955}
\definecolor{color37}{rgb}{0.659371430462219,0.674857441480988,0.920904677424953}
\definecolor{color38}{rgb}{0.703904066328539,0.66153181011268,0.921148623980379}
\definecolor{color39}{rgb}{0.742313276066846,0.644495670669399,0.918648268244515}
\definecolor{color40}{rgb}{0.779057474665895,0.625575847873961,0.915874624267216}
\definecolor{color41}{rgb}{0.815614243247047,0.603682748942063,0.912669214230223}
\definecolor{color42}{rgb}{0.853360056892316,0.577074366682852,0.908779279909119}
\definecolor{color43}{rgb}{0.893779873949281,0.542633297711204,0.903753601162158}
\definecolor{color44}{rgb}{0.904880752795055,0.543578851897922,0.868422944532749}
\definecolor{color45}{rgb}{0.907212388582589,0.55285822173336,0.831282373435747}
\definecolor{color46}{rgb}{0.909105879795907,0.560321560101729,0.798235819799659}
\definecolor{color47}{rgb}{0.910719090321778,0.566630271614636,0.76759879166572}
\definecolor{color48}{rgb}{0.912150411850359,0.572190048158244,0.738102884635612}
\definecolor{color49}{rgb}{0.913466684478731,0.577272268293219,0.708651030978847}
\definecolor{color50}{rgb}{0.914717594568828,0.5820753402252,0.678133388985302}

\begin{axis}[
axis line style={white!15!black},
tick align=outside,
x grid style={white!80!black},
xlabel={Week},
xmajorticks=false,
xmin=-0.5, xmax=50.5,
xtick style={color=white!15!black},
xticklabel style={rotate=90.0},
y grid style={white!80!black},
ylabel={Variation},
ymajorticks=false,
ymin=-0.138383838383838, ymax=3.72087542087542,
ytick style={color=white!15!black}
]
\path [draw=white!27.0588235294118!black, fill=color0, semithick]
(axis cs:-0.4,0.375)
--(axis cs:0.4,0.375)
--(axis cs:0.4,0.575)
--(axis cs:-0.4,0.575)
--(axis cs:-0.4,0.375)
--cycle;
\path [draw=white!27.0588235294118!black, fill=color1, semithick]
(axis cs:0.6,0.181818181818182)
--(axis cs:1.4,0.181818181818182)
--(axis cs:1.4,0.5)
--(axis cs:0.6,0.5)
--(axis cs:0.6,0.181818181818182)
--cycle;
\path [draw=white!27.0588235294118!black, fill=color2, semithick]
(axis cs:1.6,0.25)
--(axis cs:2.4,0.25)
--(axis cs:2.4,0.333333333333333)
--(axis cs:1.6,0.333333333333333)
--(axis cs:1.6,0.25)
--cycle;
\path [draw=white!27.0588235294118!black, fill=color3, semithick]
(axis cs:2.6,0.233333333333333)
--(axis cs:3.4,0.233333333333333)
--(axis cs:3.4,0.416666666666667)
--(axis cs:2.6,0.416666666666667)
--(axis cs:2.6,0.233333333333333)
--cycle;
\path [draw=white!27.0588235294118!black, fill=color4, semithick]
(axis cs:3.6,0.142857142857143)
--(axis cs:4.4,0.142857142857143)
--(axis cs:4.4,0.333333333333333)
--(axis cs:3.6,0.333333333333333)
--(axis cs:3.6,0.142857142857143)
--cycle;
\path [draw=white!27.0588235294118!black, fill=color5, semithick]
(axis cs:4.6,0.171428571428571)
--(axis cs:5.4,0.171428571428571)
--(axis cs:5.4,0.5)
--(axis cs:4.6,0.5)
--(axis cs:4.6,0.171428571428571)
--cycle;
\path [draw=white!27.0588235294118!black, fill=color6, semithick]
(axis cs:5.6,0.270833333333333)
--(axis cs:6.4,0.270833333333333)
--(axis cs:6.4,0.625)
--(axis cs:5.6,0.625)
--(axis cs:5.6,0.270833333333333)
--cycle;
\path [draw=white!27.0588235294118!black, fill=color7, semithick]
(axis cs:6.6,0.190625)
--(axis cs:7.4,0.190625)
--(axis cs:7.4,0.458333333333333)
--(axis cs:6.6,0.458333333333333)
--(axis cs:6.6,0.190625)
--cycle;
\path [draw=white!27.0588235294118!black, fill=color8, semithick]
(axis cs:7.6,0.1625)
--(axis cs:8.4,0.1625)
--(axis cs:8.4,0.366666666666667)
--(axis cs:7.6,0.366666666666667)
--(axis cs:7.6,0.1625)
--cycle;
\path [draw=white!27.0588235294118!black, fill=color9, semithick]
(axis cs:8.6,0.1875)
--(axis cs:9.4,0.1875)
--(axis cs:9.4,0.4)
--(axis cs:8.6,0.4)
--(axis cs:8.6,0.1875)
--cycle;
\path [draw=white!27.0588235294118!black, fill=color10, semithick]
(axis cs:9.6,0.21875)
--(axis cs:10.4,0.21875)
--(axis cs:10.4,0.40625)
--(axis cs:9.6,0.40625)
--(axis cs:9.6,0.21875)
--cycle;
\path [draw=white!27.0588235294118!black, fill=color11, semithick]
(axis cs:10.6,0.118055555555556)
--(axis cs:11.4,0.118055555555556)
--(axis cs:11.4,1.0625)
--(axis cs:10.6,1.0625)
--(axis cs:10.6,0.118055555555556)
--cycle;
\path [draw=white!27.0588235294118!black, fill=color12, semithick]
(axis cs:11.6,0.182352941176471)
--(axis cs:12.4,0.182352941176471)
--(axis cs:12.4,0.321428571428571)
--(axis cs:11.6,0.321428571428571)
--(axis cs:11.6,0.182352941176471)
--cycle;
\path [draw=white!27.0588235294118!black, fill=color13, semithick]
(axis cs:12.6,0.125)
--(axis cs:13.4,0.125)
--(axis cs:13.4,0.4)
--(axis cs:12.6,0.4)
--(axis cs:12.6,0.125)
--cycle;
\path [draw=white!27.0588235294118!black, fill=color14, semithick]
(axis cs:13.6,0.166666666666667)
--(axis cs:14.4,0.166666666666667)
--(axis cs:14.4,0.30952380952381)
--(axis cs:13.6,0.30952380952381)
--(axis cs:13.6,0.166666666666667)
--cycle;
\path [draw=white!27.0588235294118!black, fill=color15, semithick]
(axis cs:14.6,0.174019607843137)
--(axis cs:15.4,0.174019607843137)
--(axis cs:15.4,0.333333333333333)
--(axis cs:14.6,0.333333333333333)
--(axis cs:14.6,0.174019607843137)
--cycle;
\path [draw=white!27.0588235294118!black, fill=color16, semithick]
(axis cs:15.6,0.166666666666667)
--(axis cs:16.4,0.166666666666667)
--(axis cs:16.4,0.258928571428571)
--(axis cs:15.6,0.258928571428571)
--(axis cs:15.6,0.166666666666667)
--cycle;
\path [draw=white!27.0588235294118!black, fill=color17, semithick]
(axis cs:16.6,0.3125)
--(axis cs:17.4,0.3125)
--(axis cs:17.4,0.75)
--(axis cs:16.6,0.75)
--(axis cs:16.6,0.3125)
--cycle;
\path [draw=white!27.0588235294118!black, fill=color18, semithick]
(axis cs:17.6,0.191666666666667)
--(axis cs:18.4,0.191666666666667)
--(axis cs:18.4,0.3125)
--(axis cs:17.6,0.3125)
--(axis cs:17.6,0.191666666666667)
--cycle;
\path [draw=white!27.0588235294118!black, fill=color19, semithick]
(axis cs:18.6,0.0946291560102302)
--(axis cs:19.4,0.0946291560102302)
--(axis cs:19.4,0.333333333333333)
--(axis cs:18.6,0.333333333333333)
--(axis cs:18.6,0.0946291560102302)
--cycle;
\path [draw=white!27.0588235294118!black, fill=color20, semithick]
(axis cs:19.6,0.110105580693816)
--(axis cs:20.4,0.110105580693816)
--(axis cs:20.4,0.333333333333333)
--(axis cs:19.6,0.333333333333333)
--(axis cs:19.6,0.110105580693816)
--cycle;
\path [draw=white!27.0588235294118!black, fill=color21, semithick]
(axis cs:20.6,0.166666666666667)
--(axis cs:21.4,0.166666666666667)
--(axis cs:21.4,0.428571428571429)
--(axis cs:20.6,0.428571428571429)
--(axis cs:20.6,0.166666666666667)
--cycle;
\path [draw=white!27.0588235294118!black, fill=color22, semithick]
(axis cs:21.6,0.383333333333333)
--(axis cs:22.4,0.383333333333333)
--(axis cs:22.4,0.666666666666667)
--(axis cs:21.6,0.666666666666667)
--(axis cs:21.6,0.383333333333333)
--cycle;
\path [draw=white!27.0588235294118!black, fill=color23, semithick]
(axis cs:22.6,0.157142857142857)
--(axis cs:23.4,0.157142857142857)
--(axis cs:23.4,0.3125)
--(axis cs:22.6,0.3125)
--(axis cs:22.6,0.157142857142857)
--cycle;
\path [draw=white!27.0588235294118!black, fill=color24, semithick]
(axis cs:23.6,0.144642857142857)
--(axis cs:24.4,0.144642857142857)
--(axis cs:24.4,0.333333333333333)
--(axis cs:23.6,0.333333333333333)
--(axis cs:23.6,0.144642857142857)
--cycle;
\path [draw=white!27.0588235294118!black, fill=color25, semithick]
(axis cs:24.6,0.125)
--(axis cs:25.4,0.125)
--(axis cs:25.4,0.230769230769231)
--(axis cs:24.6,0.230769230769231)
--(axis cs:24.6,0.125)
--cycle;
\path [draw=white!27.0588235294118!black, fill=color26, semithick]
(axis cs:25.6,0.148214285714286)
--(axis cs:26.4,0.148214285714286)
--(axis cs:26.4,0.308333333333333)
--(axis cs:25.6,0.308333333333333)
--(axis cs:25.6,0.148214285714286)
--cycle;
\path [draw=white!27.0588235294118!black, fill=color27, semithick]
(axis cs:26.6,0.13961038961039)
--(axis cs:27.4,0.13961038961039)
--(axis cs:27.4,0.333333333333333)
--(axis cs:26.6,0.333333333333333)
--(axis cs:26.6,0.13961038961039)
--cycle;
\path [draw=white!27.0588235294118!black, fill=color28, semithick]
(axis cs:27.6,0.166666666666667)
--(axis cs:28.4,0.166666666666667)
--(axis cs:28.4,0.333333333333333)
--(axis cs:27.6,0.333333333333333)
--(axis cs:27.6,0.166666666666667)
--cycle;
\path [draw=white!27.0588235294118!black, fill=color29, semithick]
(axis cs:28.6,0.229166666666667)
--(axis cs:29.4,0.229166666666667)
--(axis cs:29.4,0.4)
--(axis cs:28.6,0.4)
--(axis cs:28.6,0.229166666666667)
--cycle;
\path [draw=white!27.0588235294118!black, fill=color30, semithick]
(axis cs:29.6,0.0922619047619048)
--(axis cs:30.4,0.0922619047619048)
--(axis cs:30.4,0.5)
--(axis cs:29.6,0.5)
--(axis cs:29.6,0.0922619047619048)
--cycle;
\path [draw=white!27.0588235294118!black, fill=color31, semithick]
(axis cs:30.6,0.2)
--(axis cs:31.4,0.2)
--(axis cs:31.4,0.326923076923077)
--(axis cs:30.6,0.326923076923077)
--(axis cs:30.6,0.2)
--cycle;
\path [draw=white!27.0588235294118!black, fill=color32, semithick]
(axis cs:31.6,0.133928571428571)
--(axis cs:32.4,0.133928571428571)
--(axis cs:32.4,0.32051282051282)
--(axis cs:31.6,0.32051282051282)
--(axis cs:31.6,0.133928571428571)
--cycle;
\path [draw=white!27.0588235294118!black, fill=color33, semithick]
(axis cs:32.6,0.15625)
--(axis cs:33.4,0.15625)
--(axis cs:33.4,0.333333333333333)
--(axis cs:32.6,0.333333333333333)
--(axis cs:32.6,0.15625)
--cycle;
\path [draw=white!27.0588235294118!black, fill=color34, semithick]
(axis cs:33.6,0.152597402597403)
--(axis cs:34.4,0.152597402597403)
--(axis cs:34.4,0.293269230769231)
--(axis cs:33.6,0.293269230769231)
--(axis cs:33.6,0.152597402597403)
--cycle;
\path [draw=white!27.0588235294118!black, fill=color35, semithick]
(axis cs:34.6,0.160714285714286)
--(axis cs:35.4,0.160714285714286)
--(axis cs:35.4,0.314285714285714)
--(axis cs:34.6,0.314285714285714)
--(axis cs:34.6,0.160714285714286)
--cycle;
\path [draw=white!27.0588235294118!black, fill=color36, semithick]
(axis cs:35.6,0.181818181818182)
--(axis cs:36.4,0.181818181818182)
--(axis cs:36.4,0.4)
--(axis cs:35.6,0.4)
--(axis cs:35.6,0.181818181818182)
--cycle;
\path [draw=white!27.0588235294118!black, fill=color37, semithick]
(axis cs:36.6,0.118055555555556)
--(axis cs:37.4,0.118055555555556)
--(axis cs:37.4,0.416666666666667)
--(axis cs:36.6,0.416666666666667)
--(axis cs:36.6,0.118055555555556)
--cycle;
\path [draw=white!27.0588235294118!black, fill=color38, semithick]
(axis cs:37.6,0.216666666666667)
--(axis cs:38.4,0.216666666666667)
--(axis cs:38.4,0.45)
--(axis cs:37.6,0.45)
--(axis cs:37.6,0.216666666666667)
--cycle;
\path [draw=white!27.0588235294118!black, fill=color39, semithick]
(axis cs:38.6,0.267045454545455)
--(axis cs:39.4,0.267045454545455)
--(axis cs:39.4,1)
--(axis cs:38.6,1)
--(axis cs:38.6,0.267045454545455)
--cycle;
\path [draw=white!27.0588235294118!black, fill=color40, semithick]
(axis cs:39.6,0.148809523809524)
--(axis cs:40.4,0.148809523809524)
--(axis cs:40.4,0.383333333333333)
--(axis cs:39.6,0.383333333333333)
--(axis cs:39.6,0.148809523809524)
--cycle;
\path [draw=white!27.0588235294118!black, fill=color41, semithick]
(axis cs:40.6,0.130952380952381)
--(axis cs:41.4,0.130952380952381)
--(axis cs:41.4,0.25)
--(axis cs:40.6,0.25)
--(axis cs:40.6,0.130952380952381)
--cycle;
\path [draw=white!27.0588235294118!black, fill=color42, semithick]
(axis cs:41.6,0.153846153846154)
--(axis cs:42.4,0.153846153846154)
--(axis cs:42.4,0.25)
--(axis cs:41.6,0.25)
--(axis cs:41.6,0.153846153846154)
--cycle;
\path [draw=white!27.0588235294118!black, fill=color43, semithick]
(axis cs:42.6,0.3)
--(axis cs:43.4,0.3)
--(axis cs:43.4,0.525)
--(axis cs:42.6,0.525)
--(axis cs:42.6,0.3)
--cycle;
\path [draw=white!27.0588235294118!black, fill=color44, semithick]
(axis cs:43.6,0.133333333333333)
--(axis cs:44.4,0.133333333333333)
--(axis cs:44.4,0.5)
--(axis cs:43.6,0.5)
--(axis cs:43.6,0.133333333333333)
--cycle;
\path [draw=white!27.0588235294118!black, fill=color45, semithick]
(axis cs:44.6,0.105555555555556)
--(axis cs:45.4,0.105555555555556)
--(axis cs:45.4,0.416666666666667)
--(axis cs:44.6,0.416666666666667)
--(axis cs:44.6,0.105555555555556)
--cycle;
\path [draw=white!27.0588235294118!black, fill=color46, semithick]
(axis cs:45.6,0.102777777777778)
--(axis cs:46.4,0.102777777777778)
--(axis cs:46.4,0.2375)
--(axis cs:45.6,0.2375)
--(axis cs:45.6,0.102777777777778)
--cycle;
\path [draw=white!27.0588235294118!black, fill=color47, semithick]
(axis cs:46.6,0.2)
--(axis cs:47.4,0.2)
--(axis cs:47.4,0.25)
--(axis cs:46.6,0.25)
--(axis cs:46.6,0.2)
--cycle;
\path [draw=white!27.0588235294118!black, fill=color48, semithick]
(axis cs:47.6,0.267857142857143)
--(axis cs:48.4,0.267857142857143)
--(axis cs:48.4,0.333333333333333)
--(axis cs:47.6,0.333333333333333)
--(axis cs:47.6,0.267857142857143)
--cycle;
\path [draw=white!27.0588235294118!black, fill=color49, semithick]
(axis cs:48.6,0.339285714285714)
--(axis cs:49.4,0.339285714285714)
--(axis cs:49.4,0.5)
--(axis cs:48.6,0.5)
--(axis cs:48.6,0.339285714285714)
--cycle;
\path [draw=white!27.0588235294118!black, fill=color50, semithick]
(axis cs:49.6,0.375)
--(axis cs:50.4,0.375)
--(axis cs:50.4,0.5)
--(axis cs:49.6,0.5)
--(axis cs:49.6,0.375)
--cycle;
\addplot [draw=blue, fill=blue, mark=*, only marks, opacity=0.8]
table{%
x  y
0 2
1 2
3 2
4 2
5 2
6 3
7 2
11 2
17 2
18 2
19 1.5
21 1.33333333333333
22 2
24 3.54545454545455
25 1.5
27 1.33333333333333
28 1.5
29 3
30 3
31 2
33 2
35 2
39 2
45 2
46 2
49 2
50 2
};
\addplot [semithick, white!27.0588235294118!black]
table {%
0 0.375
0 0.2
};
\addplot [semithick, white!27.0588235294118!black]
table {%
0 0.575
0 0.6
};
\addplot [semithick, white!27.0588235294118!black]
table {%
-0.2 0.2
0.2 0.2
};
\addplot [semithick, white!27.0588235294118!black]
table {%
-0.2 0.6
0.2 0.6
};
\addplot [black, mark=diamond*, mark size=2.5, mark options={solid,fill=white!27.0588235294118!black}, only marks]
table {%
0 2
};
\addplot [semithick, white!27.0588235294118!black]
table {%
1 0.181818181818182
1 0.111111111111111
};
\addplot [semithick, white!27.0588235294118!black]
table {%
1 0.5
1 0.5
};
\addplot [semithick, white!27.0588235294118!black]
table {%
0.8 0.111111111111111
1.2 0.111111111111111
};
\addplot [semithick, white!27.0588235294118!black]
table {%
0.8 0.5
1.2 0.5
};
\addplot [black, mark=diamond*, mark size=2.5, mark options={solid,fill=white!27.0588235294118!black}, only marks]
table {%
1 2
};
\addplot [semithick, white!27.0588235294118!black]
table {%
2 0.25
2 0.2
};
\addplot [semithick, white!27.0588235294118!black]
table {%
2 0.333333333333333
2 0.4
};
\addplot [semithick, white!27.0588235294118!black]
table {%
1.8 0.2
2.2 0.2
};
\addplot [semithick, white!27.0588235294118!black]
table {%
1.8 0.4
2.2 0.4
};
\addplot [black, mark=diamond*, mark size=2.5, mark options={solid,fill=white!27.0588235294118!black}, only marks]
table {%
2 0.5
};
\addplot [semithick, white!27.0588235294118!black]
table {%
3 0.233333333333333
3 0.2
};
\addplot [semithick, white!27.0588235294118!black]
table {%
3 0.416666666666667
3 0.5
};
\addplot [semithick, white!27.0588235294118!black]
table {%
2.8 0.2
3.2 0.2
};
\addplot [semithick, white!27.0588235294118!black]
table {%
2.8 0.5
3.2 0.5
};
\addplot [black, mark=diamond*, mark size=2.5, mark options={solid,fill=white!27.0588235294118!black}, only marks]
table {%
3 2
};
\addplot [semithick, white!27.0588235294118!black]
table {%
4 0.142857142857143
4 0.0714285714285714
};
\addplot [semithick, white!27.0588235294118!black]
table {%
4 0.333333333333333
4 0.5
};
\addplot [semithick, white!27.0588235294118!black]
table {%
3.8 0.0714285714285714
4.2 0.0714285714285714
};
\addplot [semithick, white!27.0588235294118!black]
table {%
3.8 0.5
4.2 0.5
};
\addplot [black, mark=diamond*, mark size=2.5, mark options={solid,fill=white!27.0588235294118!black}, only marks]
table {%
4 2
};
\addplot [semithick, white!27.0588235294118!black]
table {%
5 0.171428571428571
5 0.125
};
\addplot [semithick, white!27.0588235294118!black]
table {%
5 0.5
5 0.75
};
\addplot [semithick, white!27.0588235294118!black]
table {%
4.8 0.125
5.2 0.125
};
\addplot [semithick, white!27.0588235294118!black]
table {%
4.8 0.75
5.2 0.75
};
\addplot [black, mark=diamond*, mark size=2.5, mark options={solid,fill=white!27.0588235294118!black}, only marks]
table {%
5 2
};
\addplot [semithick, white!27.0588235294118!black]
table {%
6 0.270833333333333
6 0.133333333333333
};
\addplot [semithick, white!27.0588235294118!black]
table {%
6 0.625
6 0.666666666666667
};
\addplot [semithick, white!27.0588235294118!black]
table {%
5.8 0.133333333333333
6.2 0.133333333333333
};
\addplot [semithick, white!27.0588235294118!black]
table {%
5.8 0.666666666666667
6.2 0.666666666666667
};
\addplot [black, mark=diamond*, mark size=2.5, mark options={solid,fill=white!27.0588235294118!black}, only marks]
table {%
6 3
};
\addplot [semithick, white!27.0588235294118!black]
table {%
7 0.190625
7 0.142857142857143
};
\addplot [semithick, white!27.0588235294118!black]
table {%
7 0.458333333333333
7 0.5
};
\addplot [semithick, white!27.0588235294118!black]
table {%
6.8 0.142857142857143
7.2 0.142857142857143
};
\addplot [semithick, white!27.0588235294118!black]
table {%
6.8 0.5
7.2 0.5
};
\addplot [black, mark=diamond*, mark size=2.5, mark options={solid,fill=white!27.0588235294118!black}, only marks]
table {%
7 2
};
\addplot [semithick, white!27.0588235294118!black]
table {%
8 0.1625
8 0.0714285714285714
};
\addplot [semithick, white!27.0588235294118!black]
table {%
8 0.366666666666667
8 0.5
};
\addplot [semithick, white!27.0588235294118!black]
table {%
7.8 0.0714285714285714
8.2 0.0714285714285714
};
\addplot [semithick, white!27.0588235294118!black]
table {%
7.8 0.5
8.2 0.5
};
\addplot [semithick, white!27.0588235294118!black]
table {%
9 0.1875
9 0.111111111111111
};
\addplot [semithick, white!27.0588235294118!black]
table {%
9 0.4
9 0.5
};
\addplot [semithick, white!27.0588235294118!black]
table {%
8.8 0.111111111111111
9.2 0.111111111111111
};
\addplot [semithick, white!27.0588235294118!black]
table {%
8.8 0.5
9.2 0.5
};
\addplot [semithick, white!27.0588235294118!black]
table {%
10 0.21875
10 0.125
};
\addplot [semithick, white!27.0588235294118!black]
table {%
10 0.40625
10 0.5
};
\addplot [semithick, white!27.0588235294118!black]
table {%
9.8 0.125
10.2 0.125
};
\addplot [semithick, white!27.0588235294118!black]
table {%
9.8 0.5
10.2 0.5
};
\addplot [semithick, white!27.0588235294118!black]
table {%
11 0.118055555555556
11 0.111111111111111
};
\addplot [semithick, white!27.0588235294118!black]
table {%
11 1.0625
11 2
};
\addplot [semithick, white!27.0588235294118!black]
table {%
10.8 0.111111111111111
11.2 0.111111111111111
};
\addplot [semithick, white!27.0588235294118!black]
table {%
10.8 2
11.2 2
};
\addplot [semithick, white!27.0588235294118!black]
table {%
12 0.182352941176471
12 0.166666666666667
};
\addplot [semithick, white!27.0588235294118!black]
table {%
12 0.321428571428571
12 0.5
};
\addplot [semithick, white!27.0588235294118!black]
table {%
11.8 0.166666666666667
12.2 0.166666666666667
};
\addplot [semithick, white!27.0588235294118!black]
table {%
11.8 0.5
12.2 0.5
};
\addplot [semithick, white!27.0588235294118!black]
table {%
13 0.125
13 0.0909090909090909
};
\addplot [semithick, white!27.0588235294118!black]
table {%
13 0.4
13 0.6
};
\addplot [semithick, white!27.0588235294118!black]
table {%
12.8 0.0909090909090909
13.2 0.0909090909090909
};
\addplot [semithick, white!27.0588235294118!black]
table {%
12.8 0.6
13.2 0.6
};
\addplot [semithick, white!27.0588235294118!black]
table {%
14 0.166666666666667
14 0.111111111111111
};
\addplot [semithick, white!27.0588235294118!black]
table {%
14 0.30952380952381
14 0.5
};
\addplot [semithick, white!27.0588235294118!black]
table {%
13.8 0.111111111111111
14.2 0.111111111111111
};
\addplot [semithick, white!27.0588235294118!black]
table {%
13.8 0.5
14.2 0.5
};
\addplot [semithick, white!27.0588235294118!black]
table {%
15 0.174019607843137
15 0.1
};
\addplot [semithick, white!27.0588235294118!black]
table {%
15 0.333333333333333
15 0.5
};
\addplot [semithick, white!27.0588235294118!black]
table {%
14.8 0.1
15.2 0.1
};
\addplot [semithick, white!27.0588235294118!black]
table {%
14.8 0.5
15.2 0.5
};
\addplot [semithick, white!27.0588235294118!black]
table {%
16 0.166666666666667
16 0.111111111111111
};
\addplot [semithick, white!27.0588235294118!black]
table {%
16 0.258928571428571
16 0.333333333333333
};
\addplot [semithick, white!27.0588235294118!black]
table {%
15.8 0.111111111111111
16.2 0.111111111111111
};
\addplot [semithick, white!27.0588235294118!black]
table {%
15.8 0.333333333333333
16.2 0.333333333333333
};
\addplot [semithick, white!27.0588235294118!black]
table {%
17 0.3125
17 0.25
};
\addplot [semithick, white!27.0588235294118!black]
table {%
17 0.75
17 0.75
};
\addplot [semithick, white!27.0588235294118!black]
table {%
16.8 0.25
17.2 0.25
};
\addplot [semithick, white!27.0588235294118!black]
table {%
16.8 0.75
17.2 0.75
};
\addplot [black, mark=diamond*, mark size=2.5, mark options={solid,fill=white!27.0588235294118!black}, only marks]
table {%
17 2
};
\addplot [semithick, white!27.0588235294118!black]
table {%
18 0.191666666666667
18 0.0714285714285714
};
\addplot [semithick, white!27.0588235294118!black]
table {%
18 0.3125
18 0.3125
};
\addplot [semithick, white!27.0588235294118!black]
table {%
17.8 0.0714285714285714
18.2 0.0714285714285714
};
\addplot [semithick, white!27.0588235294118!black]
table {%
17.8 0.3125
18.2 0.3125
};
\addplot [black, mark=diamond*, mark size=2.5, mark options={solid,fill=white!27.0588235294118!black}, only marks]
table {%
18 2
18 0.5
};
\addplot [semithick, white!27.0588235294118!black]
table {%
19 0.0946291560102302
19 0.0555555555555556
};
\addplot [semithick, white!27.0588235294118!black]
table {%
19 0.333333333333333
19 0.545454545454545
};
\addplot [semithick, white!27.0588235294118!black]
table {%
18.8 0.0555555555555556
19.2 0.0555555555555556
};
\addplot [semithick, white!27.0588235294118!black]
table {%
18.8 0.545454545454545
19.2 0.545454545454545
};
\addplot [black, mark=diamond*, mark size=2.5, mark options={solid,fill=white!27.0588235294118!black}, only marks]
table {%
19 1.5
};
\addplot [semithick, white!27.0588235294118!black]
table {%
20 0.110105580693816
20 0.0769230769230769
};
\addplot [semithick, white!27.0588235294118!black]
table {%
20 0.333333333333333
20 0.545454545454545
};
\addplot [semithick, white!27.0588235294118!black]
table {%
19.8 0.0769230769230769
20.2 0.0769230769230769
};
\addplot [semithick, white!27.0588235294118!black]
table {%
19.8 0.545454545454545
20.2 0.545454545454545
};
\addplot [semithick, white!27.0588235294118!black]
table {%
21 0.166666666666667
21 0.111111111111111
};
\addplot [semithick, white!27.0588235294118!black]
table {%
21 0.428571428571429
21 0.5
};
\addplot [semithick, white!27.0588235294118!black]
table {%
20.8 0.111111111111111
21.2 0.111111111111111
};
\addplot [semithick, white!27.0588235294118!black]
table {%
20.8 0.5
21.2 0.5
};
\addplot [black, mark=diamond*, mark size=2.5, mark options={solid,fill=white!27.0588235294118!black}, only marks]
table {%
21 1.33333333333333
};
\addplot [semithick, white!27.0588235294118!black]
table {%
22 0.383333333333333
22 0.291666666666667
};
\addplot [semithick, white!27.0588235294118!black]
table {%
22 0.666666666666667
22 0.666666666666667
};
\addplot [semithick, white!27.0588235294118!black]
table {%
21.8 0.291666666666667
22.2 0.291666666666667
};
\addplot [semithick, white!27.0588235294118!black]
table {%
21.8 0.666666666666667
22.2 0.666666666666667
};
\addplot [black, mark=diamond*, mark size=2.5, mark options={solid,fill=white!27.0588235294118!black}, only marks]
table {%
22 2
};
\addplot [semithick, white!27.0588235294118!black]
table {%
23 0.157142857142857
23 0.0714285714285714
};
\addplot [semithick, white!27.0588235294118!black]
table {%
23 0.3125
23 0.333333333333333
};
\addplot [semithick, white!27.0588235294118!black]
table {%
22.8 0.0714285714285714
23.2 0.0714285714285714
};
\addplot [semithick, white!27.0588235294118!black]
table {%
22.8 0.333333333333333
23.2 0.333333333333333
};
\addplot [semithick, white!27.0588235294118!black]
table {%
24 0.144642857142857
24 0.0909090909090909
};
\addplot [semithick, white!27.0588235294118!black]
table {%
24 0.333333333333333
24 0.6
};
\addplot [semithick, white!27.0588235294118!black]
table {%
23.8 0.0909090909090909
24.2 0.0909090909090909
};
\addplot [semithick, white!27.0588235294118!black]
table {%
23.8 0.6
24.2 0.6
};
\addplot [black, mark=diamond*, mark size=2.5, mark options={solid,fill=white!27.0588235294118!black}, only marks]
table {%
24 3.54545454545455
};
\addplot [semithick, white!27.0588235294118!black]
table {%
25 0.125
25 0.0476190476190476
};
\addplot [semithick, white!27.0588235294118!black]
table {%
25 0.230769230769231
25 0.333333333333333
};
\addplot [semithick, white!27.0588235294118!black]
table {%
24.8 0.0476190476190476
25.2 0.0476190476190476
};
\addplot [semithick, white!27.0588235294118!black]
table {%
24.8 0.333333333333333
25.2 0.333333333333333
};
\addplot [black, mark=diamond*, mark size=2.5, mark options={solid,fill=white!27.0588235294118!black}, only marks]
table {%
25 1.5
25 0.5
};
\addplot [semithick, white!27.0588235294118!black]
table {%
26 0.148214285714286
26 0.0714285714285714
};
\addplot [semithick, white!27.0588235294118!black]
table {%
26 0.308333333333333
26 0.5
};
\addplot [semithick, white!27.0588235294118!black]
table {%
25.8 0.0714285714285714
26.2 0.0714285714285714
};
\addplot [semithick, white!27.0588235294118!black]
table {%
25.8 0.5
26.2 0.5
};
\addplot [semithick, white!27.0588235294118!black]
table {%
27 0.13961038961039
27 0.0714285714285714
};
\addplot [semithick, white!27.0588235294118!black]
table {%
27 0.333333333333333
27 0.5
};
\addplot [semithick, white!27.0588235294118!black]
table {%
26.8 0.0714285714285714
27.2 0.0714285714285714
};
\addplot [semithick, white!27.0588235294118!black]
table {%
26.8 0.5
27.2 0.5
};
\addplot [black, mark=diamond*, mark size=2.5, mark options={solid,fill=white!27.0588235294118!black}, only marks]
table {%
27 1.33333333333333
};
\addplot [semithick, white!27.0588235294118!black]
table {%
28 0.166666666666667
28 0.166666666666667
};
\addplot [semithick, white!27.0588235294118!black]
table {%
28 0.333333333333333
28 0.333333333333333
};
\addplot [semithick, white!27.0588235294118!black]
table {%
27.8 0.166666666666667
28.2 0.166666666666667
};
\addplot [semithick, white!27.0588235294118!black]
table {%
27.8 0.333333333333333
28.2 0.333333333333333
};
\addplot [black, mark=diamond*, mark size=2.5, mark options={solid,fill=white!27.0588235294118!black}, only marks]
table {%
28 1.5
};
\addplot [semithick, white!27.0588235294118!black]
table {%
29 0.229166666666667
29 0.111111111111111
};
\addplot [semithick, white!27.0588235294118!black]
table {%
29 0.4
29 0.4
};
\addplot [semithick, white!27.0588235294118!black]
table {%
28.8 0.111111111111111
29.2 0.111111111111111
};
\addplot [semithick, white!27.0588235294118!black]
table {%
28.8 0.4
29.2 0.4
};
\addplot [black, mark=diamond*, mark size=2.5, mark options={solid,fill=white!27.0588235294118!black}, only marks]
table {%
29 3
};
\addplot [semithick, white!27.0588235294118!black]
table {%
30 0.0922619047619048
30 0.037037037037037
};
\addplot [semithick, white!27.0588235294118!black]
table {%
30 0.5
30 0.75
};
\addplot [semithick, white!27.0588235294118!black]
table {%
29.8 0.037037037037037
30.2 0.037037037037037
};
\addplot [semithick, white!27.0588235294118!black]
table {%
29.8 0.75
30.2 0.75
};
\addplot [black, mark=diamond*, mark size=2.5, mark options={solid,fill=white!27.0588235294118!black}, only marks]
table {%
30 3
};
\addplot [semithick, white!27.0588235294118!black]
table {%
31 0.2
31 0.0714285714285714
};
\addplot [semithick, white!27.0588235294118!black]
table {%
31 0.326923076923077
31 0.5
};
\addplot [semithick, white!27.0588235294118!black]
table {%
30.8 0.0714285714285714
31.2 0.0714285714285714
};
\addplot [semithick, white!27.0588235294118!black]
table {%
30.8 0.5
31.2 0.5
};
\addplot [black, mark=diamond*, mark size=2.5, mark options={solid,fill=white!27.0588235294118!black}, only marks]
table {%
31 2
31 0.666666666666667
};
\addplot [semithick, white!27.0588235294118!black]
table {%
32 0.133928571428571
32 0.0714285714285714
};
\addplot [semithick, white!27.0588235294118!black]
table {%
32 0.32051282051282
32 0.6
};
\addplot [semithick, white!27.0588235294118!black]
table {%
31.8 0.0714285714285714
32.2 0.0714285714285714
};
\addplot [semithick, white!27.0588235294118!black]
table {%
31.8 0.6
32.2 0.6
};
\addplot [black, mark=diamond*, mark size=2.5, mark options={solid,fill=white!27.0588235294118!black}, only marks]
table {%
32 0.666666666666667
};
\addplot [semithick, white!27.0588235294118!black]
table {%
33 0.15625
33 0.1
};
\addplot [semithick, white!27.0588235294118!black]
table {%
33 0.333333333333333
33 0.333333333333333
};
\addplot [semithick, white!27.0588235294118!black]
table {%
32.8 0.1
33.2 0.1
};
\addplot [semithick, white!27.0588235294118!black]
table {%
32.8 0.333333333333333
33.2 0.333333333333333
};
\addplot [black, mark=diamond*, mark size=2.5, mark options={solid,fill=white!27.0588235294118!black}, only marks]
table {%
33 2
33 0.7
};
\addplot [semithick, white!27.0588235294118!black]
table {%
34 0.152597402597403
34 0.0714285714285714
};
\addplot [semithick, white!27.0588235294118!black]
table {%
34 0.293269230769231
34 0.5
};
\addplot [semithick, white!27.0588235294118!black]
table {%
33.8 0.0714285714285714
34.2 0.0714285714285714
};
\addplot [semithick, white!27.0588235294118!black]
table {%
33.8 0.5
34.2 0.5
};
\addplot [black, mark=diamond*, mark size=2.5, mark options={solid,fill=white!27.0588235294118!black}, only marks]
table {%
34 0.666666666666667
34 0.6
};
\addplot [semithick, white!27.0588235294118!black]
table {%
35 0.160714285714286
35 0.0740740740740741
};
\addplot [semithick, white!27.0588235294118!black]
table {%
35 0.314285714285714
35 0.428571428571429
};
\addplot [semithick, white!27.0588235294118!black]
table {%
34.8 0.0740740740740741
35.2 0.0740740740740741
};
\addplot [semithick, white!27.0588235294118!black]
table {%
34.8 0.428571428571429
35.2 0.428571428571429
};
\addplot [black, mark=diamond*, mark size=2.5, mark options={solid,fill=white!27.0588235294118!black}, only marks]
table {%
35 2
};
\addplot [semithick, white!27.0588235294118!black]
table {%
36 0.181818181818182
36 0.0588235294117647
};
\addplot [semithick, white!27.0588235294118!black]
table {%
36 0.4
36 0.666666666666667
};
\addplot [semithick, white!27.0588235294118!black]
table {%
35.8 0.0588235294117647
36.2 0.0588235294117647
};
\addplot [semithick, white!27.0588235294118!black]
table {%
35.8 0.666666666666667
36.2 0.666666666666667
};
\addplot [semithick, white!27.0588235294118!black]
table {%
37 0.118055555555556
37 0.0714285714285714
};
\addplot [semithick, white!27.0588235294118!black]
table {%
37 0.416666666666667
37 0.6
};
\addplot [semithick, white!27.0588235294118!black]
table {%
36.8 0.0714285714285714
37.2 0.0714285714285714
};
\addplot [semithick, white!27.0588235294118!black]
table {%
36.8 0.6
37.2 0.6
};
\addplot [semithick, white!27.0588235294118!black]
table {%
38 0.216666666666667
38 0.1
};
\addplot [semithick, white!27.0588235294118!black]
table {%
38 0.45
38 0.666666666666667
};
\addplot [semithick, white!27.0588235294118!black]
table {%
37.8 0.1
38.2 0.1
};
\addplot [semithick, white!27.0588235294118!black]
table {%
37.8 0.666666666666667
38.2 0.666666666666667
};
\addplot [semithick, white!27.0588235294118!black]
table {%
39 0.267045454545455
39 0.25
};
\addplot [semithick, white!27.0588235294118!black]
table {%
39 1
39 2
};
\addplot [semithick, white!27.0588235294118!black]
table {%
38.8 0.25
39.2 0.25
};
\addplot [semithick, white!27.0588235294118!black]
table {%
38.8 2
39.2 2
};
\addplot [semithick, white!27.0588235294118!black]
table {%
40 0.148809523809524
40 0.0869565217391304
};
\addplot [semithick, white!27.0588235294118!black]
table {%
40 0.383333333333333
40 0.666666666666667
};
\addplot [semithick, white!27.0588235294118!black]
table {%
39.8 0.0869565217391304
40.2 0.0869565217391304
};
\addplot [semithick, white!27.0588235294118!black]
table {%
39.8 0.666666666666667
40.2 0.666666666666667
};
\addplot [semithick, white!27.0588235294118!black]
table {%
41 0.130952380952381
41 0.0526315789473684
};
\addplot [semithick, white!27.0588235294118!black]
table {%
41 0.25
41 0.363636363636364
};
\addplot [semithick, white!27.0588235294118!black]
table {%
40.8 0.0526315789473684
41.2 0.0526315789473684
};
\addplot [semithick, white!27.0588235294118!black]
table {%
40.8 0.363636363636364
41.2 0.363636363636364
};
\addplot [semithick, white!27.0588235294118!black]
table {%
42 0.153846153846154
42 0.1
};
\addplot [semithick, white!27.0588235294118!black]
table {%
42 0.25
42 0.25
};
\addplot [semithick, white!27.0588235294118!black]
table {%
41.8 0.1
42.2 0.1
};
\addplot [semithick, white!27.0588235294118!black]
table {%
41.8 0.25
42.2 0.25
};
\addplot [black, mark=diamond*, mark size=2.5, mark options={solid,fill=white!27.0588235294118!black}, only marks]
table {%
42 0.666666666666667
42 0.5
};
\addplot [semithick, white!27.0588235294118!black]
table {%
43 0.3
43 0.2
};
\addplot [semithick, white!27.0588235294118!black]
table {%
43 0.525
43 0.6
};
\addplot [semithick, white!27.0588235294118!black]
table {%
42.8 0.2
43.2 0.2
};
\addplot [semithick, white!27.0588235294118!black]
table {%
42.8 0.6
43.2 0.6
};
\addplot [semithick, white!27.0588235294118!black]
table {%
44 0.133333333333333
44 0.0769230769230769
};
\addplot [semithick, white!27.0588235294118!black]
table {%
44 0.5
44 0.6
};
\addplot [semithick, white!27.0588235294118!black]
table {%
43.8 0.0769230769230769
44.2 0.0769230769230769
};
\addplot [semithick, white!27.0588235294118!black]
table {%
43.8 0.6
44.2 0.6
};
\addplot [semithick, white!27.0588235294118!black]
table {%
45 0.105555555555556
45 0.0769230769230769
};
\addplot [semithick, white!27.0588235294118!black]
table {%
45 0.416666666666667
45 0.5
};
\addplot [semithick, white!27.0588235294118!black]
table {%
44.8 0.0769230769230769
45.2 0.0769230769230769
};
\addplot [semithick, white!27.0588235294118!black]
table {%
44.8 0.5
45.2 0.5
};
\addplot [black, mark=diamond*, mark size=2.5, mark options={solid,fill=white!27.0588235294118!black}, only marks]
table {%
45 2
};
\addplot [semithick, white!27.0588235294118!black]
table {%
46 0.102777777777778
46 0.0454545454545455
};
\addplot [semithick, white!27.0588235294118!black]
table {%
46 0.2375
46 0.25
};
\addplot [semithick, white!27.0588235294118!black]
table {%
45.8 0.0454545454545455
46.2 0.0454545454545455
};
\addplot [semithick, white!27.0588235294118!black]
table {%
45.8 0.25
46.2 0.25
};
\addplot [black, mark=diamond*, mark size=2.5, mark options={solid,fill=white!27.0588235294118!black}, only marks]
table {%
46 2
46 0.5
};
\addplot [semithick, white!27.0588235294118!black]
table {%
47 0.2
47 0.2
};
\addplot [semithick, white!27.0588235294118!black]
table {%
47 0.25
47 0.25
};
\addplot [semithick, white!27.0588235294118!black]
table {%
46.8 0.2
47.2 0.2
};
\addplot [semithick, white!27.0588235294118!black]
table {%
46.8 0.25
47.2 0.25
};
\addplot [black, mark=diamond*, mark size=2.5, mark options={solid,fill=white!27.0588235294118!black}, only marks]
table {%
47 0.04
47 0.5
};
\addplot [semithick, white!27.0588235294118!black]
table {%
48 0.267857142857143
48 0.25
};
\addplot [semithick, white!27.0588235294118!black]
table {%
48 0.333333333333333
48 0.333333333333333
};
\addplot [semithick, white!27.0588235294118!black]
table {%
47.8 0.25
48.2 0.25
};
\addplot [semithick, white!27.0588235294118!black]
table {%
47.8 0.333333333333333
48.2 0.333333333333333
};
\addplot [black, mark=diamond*, mark size=2.5, mark options={solid,fill=white!27.0588235294118!black}, only marks]
table {%
48 0.111111111111111
48 0.5
};
\addplot [semithick, white!27.0588235294118!black]
table {%
49 0.339285714285714
49 0.2
};
\addplot [semithick, white!27.0588235294118!black]
table {%
49 0.5
49 0.5
};
\addplot [semithick, white!27.0588235294118!black]
table {%
48.8 0.2
49.2 0.2
};
\addplot [semithick, white!27.0588235294118!black]
table {%
48.8 0.5
49.2 0.5
};
\addplot [black, mark=diamond*, mark size=2.5, mark options={solid,fill=white!27.0588235294118!black}, only marks]
table {%
49 2
};
\addplot [semithick, white!27.0588235294118!black]
table {%
50 0.375
50 0.2
};
\addplot [semithick, white!27.0588235294118!black]
table {%
50 0.5
50 0.5
};
\addplot [semithick, white!27.0588235294118!black]
table {%
49.8 0.2
50.2 0.2
};
\addplot [semithick, white!27.0588235294118!black]
table {%
49.8 0.5
50.2 0.5
};
\addplot [black, mark=diamond*, mark size=2.5, mark options={solid,fill=white!27.0588235294118!black}, only marks]
table {%
50 2
};
\addplot [semithick, white!27.0588235294118!black]
table {%
-0.4 0.5
0.4 0.5
};
\addplot [semithick, white!27.0588235294118!black]
table {%
0.6 0.333333333333333
1.4 0.333333333333333
};
\addplot [semithick, white!27.0588235294118!black]
table {%
1.6 0.266666666666667
2.4 0.266666666666667
};
\addplot [semithick, white!27.0588235294118!black]
table {%
2.6 0.333333333333333
3.4 0.333333333333333
};
\addplot [semithick, white!27.0588235294118!black]
table {%
3.6 0.214285714285714
4.4 0.214285714285714
};
\addplot [semithick, white!27.0588235294118!black]
table {%
4.6 0.25
5.4 0.25
};
\addplot [semithick, white!27.0588235294118!black]
table {%
5.6 0.416666666666667
6.4 0.416666666666667
};
\addplot [semithick, white!27.0588235294118!black]
table {%
6.6 0.291666666666667
7.4 0.291666666666667
};
\addplot [semithick, white!27.0588235294118!black]
table {%
7.6 0.25
8.4 0.25
};
\addplot [semithick, white!27.0588235294118!black]
table {%
8.6 0.25
9.4 0.25
};
\addplot [semithick, white!27.0588235294118!black]
table {%
9.6 0.3125
10.4 0.3125
};
\addplot [semithick, white!27.0588235294118!black]
table {%
10.6 0.125
11.4 0.125
};
\addplot [semithick, white!27.0588235294118!black]
table {%
11.6 0.25
12.4 0.25
};
\addplot [semithick, white!27.0588235294118!black]
table {%
12.6 0.25
13.4 0.25
};
\addplot [semithick, white!27.0588235294118!black]
table {%
13.6 0.21875
14.4 0.21875
};
\addplot [semithick, white!27.0588235294118!black]
table {%
14.6 0.258333333333333
15.4 0.258333333333333
};
\addplot [semithick, white!27.0588235294118!black]
table {%
15.6 0.209375
16.4 0.209375
};
\addplot [semithick, white!27.0588235294118!black]
table {%
16.6 0.333333333333333
17.4 0.333333333333333
};
\addplot [semithick, white!27.0588235294118!black]
table {%
17.6 0.207142857142857
18.4 0.207142857142857
};
\addplot [semithick, white!27.0588235294118!black]
table {%
18.6 0.2
19.4 0.2
};
\addplot [semithick, white!27.0588235294118!black]
table {%
19.6 0.214285714285714
20.4 0.214285714285714
};
\addplot [semithick, white!27.0588235294118!black]
table {%
20.6 0.291666666666667
21.4 0.291666666666667
};
\addplot [semithick, white!27.0588235294118!black]
table {%
21.6 0.464285714285714
22.4 0.464285714285714
};
\addplot [semithick, white!27.0588235294118!black]
table {%
22.6 0.225
23.4 0.225
};
\addplot [semithick, white!27.0588235294118!black]
table {%
23.6 0.299107142857143
24.4 0.299107142857143
};
\addplot [semithick, white!27.0588235294118!black]
table {%
24.6 0.178571428571429
25.4 0.178571428571429
};
\addplot [semithick, white!27.0588235294118!black]
table {%
25.6 0.225
26.4 0.225
};
\addplot [semithick, white!27.0588235294118!black]
table {%
26.6 0.2
27.4 0.2
};
\addplot [semithick, white!27.0588235294118!black]
table {%
27.6 0.166666666666667
28.4 0.166666666666667
};
\addplot [semithick, white!27.0588235294118!black]
table {%
28.6 0.333333333333333
29.4 0.333333333333333
};
\addplot [semithick, white!27.0588235294118!black]
table {%
29.6 0.225
30.4 0.225
};
\addplot [semithick, white!27.0588235294118!black]
table {%
30.6 0.244047619047619
31.4 0.244047619047619
};
\addplot [semithick, white!27.0588235294118!black]
table {%
31.6 0.208333333333333
32.4 0.208333333333333
};
\addplot [semithick, white!27.0588235294118!black]
table {%
32.6 0.30952380952381
33.4 0.30952380952381
};
\addplot [semithick, white!27.0588235294118!black]
table {%
33.6 0.218452380952381
34.4 0.218452380952381
};
\addplot [semithick, white!27.0588235294118!black]
table {%
34.6 0.236111111111111
35.4 0.236111111111111
};
\addplot [semithick, white!27.0588235294118!black]
table {%
35.6 0.2
36.4 0.2
};
\addplot [semithick, white!27.0588235294118!black]
table {%
36.6 0.25
37.4 0.25
};
\addplot [semithick, white!27.0588235294118!black]
table {%
37.6 0.333333333333333
38.4 0.333333333333333
};
\addplot [semithick, white!27.0588235294118!black]
table {%
38.6 0.46969696969697
39.4 0.46969696969697
};
\addplot [semithick, white!27.0588235294118!black]
table {%
39.6 0.276190476190476
40.4 0.276190476190476
};
\addplot [semithick, white!27.0588235294118!black]
table {%
40.6 0.225
41.4 0.225
};
\addplot [semithick, white!27.0588235294118!black]
table {%
41.6 0.25
42.4 0.25
};
\addplot [semithick, white!27.0588235294118!black]
table {%
42.6 0.416666666666667
43.4 0.416666666666667
};
\addplot [semithick, white!27.0588235294118!black]
table {%
43.6 0.166666666666667
44.4 0.166666666666667
};
\addplot [semithick, white!27.0588235294118!black]
table {%
44.6 0.230769230769231
45.4 0.230769230769231
};
\addplot [semithick, white!27.0588235294118!black]
table {%
45.6 0.181878306878307
46.4 0.181878306878307
};
\addplot [semithick, white!27.0588235294118!black]
table {%
46.6 0.2
47.4 0.2
};
\addplot [semithick, white!27.0588235294118!black]
table {%
47.6 0.333333333333333
48.4 0.333333333333333
};
\addplot [semithick, white!27.0588235294118!black]
table {%
48.6 0.5
49.4 0.5
};
\addplot [semithick, white!27.0588235294118!black]
table {%
49.6 0.5
50.4 0.5
};
\end{axis}

\end{tikzpicture}

%% file: Fransquinietalrevised.bbl
\begin{thebibliography}{37}
\expandafter\ifx\csname natexlab\endcsname\relax\def\natexlab#1{#1}\fi
\providecommand{\url}[1]{\texttt{#1}}
\providecommand{\href}[2]{#2}
\providecommand{\path}[1]{#1}
\providecommand{\DOIprefix}{doi:}
\providecommand{\ArXivprefix}{arXiv:}
\providecommand{\URLprefix}{URL: }
\providecommand{\Pubmedprefix}{pmid:}
\providecommand{\doi}[1]{\href{http://dx.doi.org/#1}{\path{#1}}}
\providecommand{\Pubmed}[1]{\href{pmid:#1}{\path{#1}}}
\providecommand{\bibinfo}[2]{#2}
\ifx\xfnm\relax \def\xfnm[#1]{\unskip,\space#1}\fi
\bibitem[{Aggarwal(2013)}]{Aggarwal2013}
\bibinfo{author}{Aggarwal, C.C.}, \bibinfo{year}{2013}.
\newblock \bibinfo{title}{Outlier Analysis}.
\newblock \bibinfo{publisher}{Springer}.
\newblock \DOIprefix\doi{https://doi.org/doi.org/10.1007/978-1-4614-6396-2}.
\bibitem[{Ahmed et~al.(2017)Ahmed, Palleti and Mathur}]{ahmed2017wadi}
\bibinfo{author}{Ahmed, C.M.}, \bibinfo{author}{Palleti, V.R.},
  \bibinfo{author}{Mathur, A.P.}, \bibinfo{year}{2017}.
\newblock \bibinfo{title}{Wadi: a water distribution testbed for research in
  the design of secure cyber physical systems}, in:
  \bibinfo{booktitle}{Proceedings of the 3rd International Workshop on
  Cyber-Physical Systems for Smart Water Networks}, pp.
  \bibinfo{pages}{25--28}.
\bibitem[{Akoglu et~al.(2015)Akoglu, Tong and Koutra}]{Akoglu2015GraphSurvey}
\bibinfo{author}{Akoglu, L.}, \bibinfo{author}{Tong, H.},
  \bibinfo{author}{Koutra, D.}, \bibinfo{year}{2015}.
\newblock \bibinfo{title}{Graph based anomaly detection and description: a
  survey}.
\newblock \bibinfo{journal}{Data Mining and Knowledge Discovery}
  \bibinfo{volume}{29}, \bibinfo{pages}{626--688}.
\bibitem[{Angiulli and Pizzuti(2002)}]{angiulli2002fast}
\bibinfo{author}{Angiulli, F.}, \bibinfo{author}{Pizzuti, C.},
  \bibinfo{year}{2002}.
\newblock \bibinfo{title}{Fast outlier detection in high dimensional spaces},
  in: \bibinfo{booktitle}{European conference on principles of data mining and
  knowledge discovery}, \bibinfo{organization}{Springer}. pp.
  \bibinfo{pages}{15--27}.
\bibitem[{Blondel et~al.(2008)Blondel, Guillaume, Lambiotte and
  Lefebvre}]{Blondel2008FastNetworks}
\bibinfo{author}{Blondel, V.D.}, \bibinfo{author}{Guillaume, J.L.},
  \bibinfo{author}{Lambiotte, R.}, \bibinfo{author}{Lefebvre, E.},
  \bibinfo{year}{2008}.
\newblock \bibinfo{title}{Fast unfolding of communities in large networks}.
\newblock \bibinfo{journal}{Journal of Statistical Mechanics: Theory and
  Experiment} \bibinfo{volume}{2008}, \bibinfo{pages}{P10008}.
\bibitem[{Buades et~al.(2005)Buades, Coll and Morel}]{Buades2005a}
\bibinfo{author}{Buades, A.}, \bibinfo{author}{Coll, B.},
  \bibinfo{author}{Morel, J.M.}, \bibinfo{year}{2005}.
\newblock \bibinfo{title}{A review of image denoising algorithms, with a new
  one}.
\newblock \bibinfo{journal}{Multiscale Modeling \& Simulation}
  \bibinfo{volume}{4}, \bibinfo{pages}{490--530}.
\bibitem[{Chen et~al.(2014)Chen, Sandryhaila, Moura and Kovacevic}]{Chen2014}
\bibinfo{author}{Chen, S.}, \bibinfo{author}{Sandryhaila, A.},
  \bibinfo{author}{Moura, J.M.}, \bibinfo{author}{Kovacevic, J.},
  \bibinfo{year}{2014}.
\newblock \bibinfo{title}{Signal denoising on graphs via graph filtering}, in:
  \bibinfo{booktitle}{2014 IEEE Global Conference on Signal and Information
  Processing (GlobalSIP)}, \bibinfo{organization}{IEEE}. pp.
  \bibinfo{pages}{872--876}.
\bibitem[{Chen et~al.(2021)Chen, Chen, Zhang, Yuan and
  Cheng}]{chen2021learning}
\bibinfo{author}{Chen, Z.}, \bibinfo{author}{Chen, D.}, \bibinfo{author}{Zhang,
  X.}, \bibinfo{author}{Yuan, Z.}, \bibinfo{author}{Cheng, X.},
  \bibinfo{year}{2021}.
\newblock \bibinfo{title}{Learning graph structures with transformer for
  multivariate time series anomaly detection in {IoT}}.
\newblock \bibinfo{journal}{IEEE Internet of Things Journal} .
\bibitem[{Choi et~al.(2021)Choi, Yi, Park and Yoon}]{choi2021deep}
\bibinfo{author}{Choi, K.}, \bibinfo{author}{Yi, J.}, \bibinfo{author}{Park,
  C.}, \bibinfo{author}{Yoon, S.}, \bibinfo{year}{2021}.
\newblock \bibinfo{title}{Deep learning for anomaly detection in time-series
  data: Review, analysis, and guidelines}.
\newblock \bibinfo{journal}{IEEE Access} .
\bibitem[{Chung and Graham(1997)}]{ChungFanRKandGraham1997}
\bibinfo{author}{Chung, F.R.}, \bibinfo{author}{Graham, F.C.},
  \bibinfo{year}{1997}.
\newblock \bibinfo{title}{Spectral graph theory}.
\newblock \bibinfo{publisher}{American Mathematical Soc.}
\bibitem[{Dem{\v{s}}ar(2006)}]{demvsar2006statistical}
\bibinfo{author}{Dem{\v{s}}ar, J.}, \bibinfo{year}{2006}.
\newblock \bibinfo{title}{Statistical comparisons of classifiers over multiple
  data sets}.
\newblock \bibinfo{journal}{The Journal of Machine Learning Research}
  \bibinfo{volume}{7}, \bibinfo{pages}{1--30}.
\bibitem[{Deng and Hooi(2021)}]{deng2021graph}
\bibinfo{author}{Deng, A.}, \bibinfo{author}{Hooi, B.}, \bibinfo{year}{2021}.
\newblock \bibinfo{title}{Graph neural network-based anomaly detection in
  multivariate time series}, in: \bibinfo{booktitle}{Proceedings of the AAAI
  Conference on Artificial Intelligence}, pp. \bibinfo{pages}{4027--4035}.
\bibitem[{Dong et~al.(2019)Dong, Thanou, Rabbat and
  Frossard}]{dong2019learning}
\bibinfo{author}{Dong, X.}, \bibinfo{author}{Thanou, D.},
  \bibinfo{author}{Rabbat, M.}, \bibinfo{author}{Frossard, P.},
  \bibinfo{year}{2019}.
\newblock \bibinfo{title}{Learning graphs from data: A signal representation
  perspective}.
\newblock \bibinfo{journal}{IEEE Signal Processing Magazine}
  \bibinfo{volume}{36}, \bibinfo{pages}{44--63}.
\bibitem[{Egilmez and Ortega(2014)}]{Qualcomm2014SpectralGraphs}
\bibinfo{author}{Egilmez, H.E.}, \bibinfo{author}{Ortega, A.},
  \bibinfo{year}{2014}.
\newblock \bibinfo{title}{Spectral anomaly detection using graph-based
  filtering for wireless sensor networks}, in: \bibinfo{booktitle}{2014 IEEE
  International Conference on Acoustics, Speech and Signal Processing
  (ICASSP)}, \bibinfo{organization}{IEEE}. pp. \bibinfo{pages}{1085--1089}.
\bibitem[{Francisquini et~al.(2021)Francisquini, Berton, Soares, Pessotti,
  Camacho, Andrade-Silva, Barcick, Serrano, Chammas, Nascimento and
  Zelanis}]{francisquini2021community}
\bibinfo{author}{Francisquini, R.}, \bibinfo{author}{Berton, R.},
  \bibinfo{author}{Soares, S.G.}, \bibinfo{author}{Pessotti, D.S.},
  \bibinfo{author}{Camacho, M.F.}, \bibinfo{author}{Andrade-Silva, D.},
  \bibinfo{author}{Barcick, U.}, \bibinfo{author}{Serrano, S.M.},
  \bibinfo{author}{Chammas, R.}, \bibinfo{author}{Nascimento, M.C.V.},
  \bibinfo{author}{Zelanis, A.}, \bibinfo{year}{2021}.
\newblock \bibinfo{title}{Community-based network analyses reveal emerging
  connectivity patterns of protein-protein interactions in murine melanoma
  secretome}.
\newblock \bibinfo{journal}{Journal of Proteomics} \bibinfo{volume}{232},
  \bibinfo{pages}{104063}.
\bibitem[{Gao et~al.(2010)Gao, Liang, Fan, Wang, Sun and Han}]{GaoOn}
\bibinfo{author}{Gao, J.}, \bibinfo{author}{Liang, F.}, \bibinfo{author}{Fan,
  W.}, \bibinfo{author}{Wang, C.}, \bibinfo{author}{Sun, Y.},
  \bibinfo{author}{Han, J.}, \bibinfo{year}{2010}.
\newblock \bibinfo{title}{On community outliers and their efficient detection
  in information networks}, in: \bibinfo{booktitle}{Proceedings of the 16th ACM
  SIGKDD International Conference on Knowledge Discovery and Data Mining}, pp.
  \bibinfo{pages}{813--822}.
\bibitem[{Gao and Yu(2021)}]{gao2021fault}
\bibinfo{author}{Gao, Y.}, \bibinfo{author}{Yu, D.}, \bibinfo{year}{2021}.
\newblock \bibinfo{title}{Fault diagnosis of rolling bearing based on laplacian
  regularization}.
\newblock \bibinfo{journal}{Applied Soft Computing} \bibinfo{volume}{111},
  \bibinfo{pages}{107651}.
\bibitem[{Lancichinetti and Fortunato(2009)}]{Lancichinetti2009}
\bibinfo{author}{Lancichinetti, A.}, \bibinfo{author}{Fortunato, S.},
  \bibinfo{year}{2009}.
\newblock \bibinfo{title}{Community detection algorithms: a comparative
  analysis}.
\newblock \bibinfo{journal}{Physical Review E} \bibinfo{volume}{80},
  \bibinfo{pages}{056117}.
\bibitem[{Lazarevic and Kumar(2005)}]{lazarevic2005feature}
\bibinfo{author}{Lazarevic, A.}, \bibinfo{author}{Kumar, V.},
  \bibinfo{year}{2005}.
\newblock \bibinfo{title}{Feature bagging for outlier detection}, in:
  \bibinfo{booktitle}{Proceedings of the eleventh ACM SIGKDD international
  conference on Knowledge discovery in data mining}, pp.
  \bibinfo{pages}{157--166}.
\bibitem[{Li et~al.(2019)Li, Chen, Jin, Shi, Goh and Ng}]{li2019mad}
\bibinfo{author}{Li, D.}, \bibinfo{author}{Chen, D.}, \bibinfo{author}{Jin,
  B.}, \bibinfo{author}{Shi, L.}, \bibinfo{author}{Goh, J.},
  \bibinfo{author}{Ng, S.K.}, \bibinfo{year}{2019}.
\newblock \bibinfo{title}{Mad-gan: Multivariate anomaly detection for time
  series data with generative adversarial networks}, in:
  \bibinfo{booktitle}{International Conference on Artificial Neural Networks},
  \bibinfo{organization}{Springer}. pp. \bibinfo{pages}{703--716}.
\bibitem[{Li et~al.(2021)Li, Izakian, Pedrycz and Jamal}]{li2021clustering}
\bibinfo{author}{Li, J.}, \bibinfo{author}{Izakian, H.},
  \bibinfo{author}{Pedrycz, W.}, \bibinfo{author}{Jamal, I.},
  \bibinfo{year}{2021}.
\newblock \bibinfo{title}{Clustering-based anomaly detection in multivariate
  time series data}.
\newblock \bibinfo{journal}{Applied Soft Computing} \bibinfo{volume}{100},
  \bibinfo{pages}{106919}.
\bibitem[{Ma et~al.(2021)Ma, Wu, Xue, Yang, Zhou, Sheng, Xiong and
  Akoglu}]{ma2021comprehensive}
\bibinfo{author}{Ma, X.}, \bibinfo{author}{Wu, J.}, \bibinfo{author}{Xue, S.},
  \bibinfo{author}{Yang, J.}, \bibinfo{author}{Zhou, C.},
  \bibinfo{author}{Sheng, Q.Z.}, \bibinfo{author}{Xiong, H.},
  \bibinfo{author}{Akoglu, L.}, \bibinfo{year}{2021}.
\newblock \bibinfo{title}{A comprehensive survey on graph anomaly detection
  with deep learning}.
\newblock \bibinfo{journal}{IEEE Transactions on Knowledge and Data
  Engineering} .
\bibitem[{Mathur and Tippenhauer(2016)}]{mathur2016swat}
\bibinfo{author}{Mathur, A.P.}, \bibinfo{author}{Tippenhauer, N.O.},
  \bibinfo{year}{2016}.
\newblock \bibinfo{title}{Swat: A water treatment testbed for research and
  training on ics security}, in: \bibinfo{booktitle}{2016 international
  workshop on cyber-physical systems for smart water networks (CySWater)},
  \bibinfo{organization}{IEEE}. pp. \bibinfo{pages}{31--36}.
\bibitem[{Meert et~al.(2020)Meert, Hendrickx and
  Craenendonck}]{wannes_meert_2020_3981067}
\bibinfo{author}{Meert, W.}, \bibinfo{author}{Hendrickx, K.},
  \bibinfo{author}{Craenendonck, T.V.}, \bibinfo{year}{2020}.
\newblock \bibinfo{title}{wannesm/dtaidistance v2.0.0}.
\newblock \DOIprefix\doi{10.5281/zenodo.3981067}.
\bibitem[{M{\"u}ller et~al.(2013)M{\"u}ller, S{\'a}nchez, M{\"u}lle and
  B{\"o}hm}]{Muller2013}
\bibinfo{author}{M{\"u}ller, E.}, \bibinfo{author}{S{\'a}nchez, P.I.},
  \bibinfo{author}{M{\"u}lle, Y.}, \bibinfo{author}{B{\"o}hm, K.},
  \bibinfo{year}{2013}.
\newblock \bibinfo{title}{Ranking outlier nodes in subspaces of attributed
  graphs}, in: \bibinfo{booktitle}{2013 IEEE 29th International Conference on
  Data Engineering Workshops (ICDEW)}, \bibinfo{organization}{IEEE}. pp.
  \bibinfo{pages}{216--222}.
\bibitem[{Newman and Girvan(2004)}]{Newman2004FindingNetworks}
\bibinfo{author}{Newman, M.E.}, \bibinfo{author}{Girvan, M.},
  \bibinfo{year}{2004}.
\newblock \bibinfo{title}{Finding and evaluating community structure in
  networks}.
\newblock \bibinfo{journal}{Physical Review E} \bibinfo{volume}{69},
  \bibinfo{pages}{026113}.
\bibitem[{Ortega et~al.(2018)Ortega, Frossard, Kova{\v{c}}evi{\'c}, Moura and
  Vandergheynst}]{ortega2018graph}
\bibinfo{author}{Ortega, A.}, \bibinfo{author}{Frossard, P.},
  \bibinfo{author}{Kova{\v{c}}evi{\'c}, J.}, \bibinfo{author}{Moura, J.M.},
  \bibinfo{author}{Vandergheynst, P.}, \bibinfo{year}{2018}.
\newblock \bibinfo{title}{Graph signal processing: Overview, challenges, and
  applications}.
\newblock \bibinfo{journal}{Proceedings of the IEEE} \bibinfo{volume}{106},
  \bibinfo{pages}{808--828}.
\bibitem[{Park et~al.(2018)Park, Hoshi and Kemp}]{park2018multimodal}
\bibinfo{author}{Park, D.}, \bibinfo{author}{Hoshi, Y.}, \bibinfo{author}{Kemp,
  C.C.}, \bibinfo{year}{2018}.
\newblock \bibinfo{title}{A multimodal anomaly detector for robot-assisted
  feeding using an lstm-based variational autoencoder}.
\newblock \bibinfo{journal}{IEEE Robotics and Automation Letters}
  \bibinfo{volume}{3}, \bibinfo{pages}{1544--1551}.
\bibitem[{P{\"{u}}schel and Moura(2008a)}]{Puschel2008a}
\bibinfo{author}{P{\"{u}}schel, M.}, \bibinfo{author}{Moura, J.M.},
  \bibinfo{year}{2008}a.
\newblock \bibinfo{title}{{Algebraic signal processing theory: 1-D space}}.
\newblock \bibinfo{journal}{IEEE Transactions on Signal Processing}
  \bibinfo{volume}{56}, \bibinfo{pages}{3586--3599}.
\bibitem[{P{\"{u}}schel and Moura(2008b)}]{Puschel2008}
\bibinfo{author}{P{\"{u}}schel, P.}, \bibinfo{author}{Moura, J.M.},
  \bibinfo{year}{2008}b.
\newblock \bibinfo{title}{{Algebraic signal processing theory: Foundation and
  1-D time}}.
\newblock \bibinfo{journal}{IEEE Transactions on Signal Processing}
  \bibinfo{volume}{56}, \bibinfo{pages}{3572--3585}.
\bibitem[{Saito and Rehmsmeier(2015)}]{Saito2015TheDatasets}
\bibinfo{author}{Saito, T.}, \bibinfo{author}{Rehmsmeier, M.},
  \bibinfo{year}{2015}.
\newblock \bibinfo{title}{The precision-recall plot is more informative than
  the roc plot when evaluating binary classifiers on imbalanced datasets}.
\newblock \bibinfo{journal}{PloS One} \bibinfo{volume}{10},
  \bibinfo{pages}{e0118432}.
\bibitem[{Sakoe(1971)}]{sakoe1971dynamic}
\bibinfo{author}{Sakoe, H.}, \bibinfo{year}{1971}.
\newblock \bibinfo{title}{Dynamic-programming approach to continuous speech
  recognition}, in: \bibinfo{booktitle}{1971 Proc. the International Congress
  of Acoustics, Budapest}.
\bibitem[{Sandryhaila and Moura(2013)}]{AliakseiSandryhaila2013}
\bibinfo{author}{Sandryhaila, A.}, \bibinfo{author}{Moura, J.M.},
  \bibinfo{year}{2013}.
\newblock \bibinfo{title}{Discrete signal processing on graphs: Graph fourier
  transform}, in: \bibinfo{booktitle}{2013 IEEE International Conference on
  Acoustics, Speech and Signal Processing}, \bibinfo{organization}{IEEE}. pp.
  \bibinfo{pages}{6167--6170}.
\bibitem[{Sandryhaila and Moura(2014)}]{Sandryhaila2014}
\bibinfo{author}{Sandryhaila, A.}, \bibinfo{author}{Moura, J.M.},
  \bibinfo{year}{2014}.
\newblock \bibinfo{title}{Discrete signal processing on graphs: Frequency
  analysis}.
\newblock \bibinfo{journal}{IEEE Transactions on Signal Processing}
  \bibinfo{volume}{62}, \bibinfo{pages}{3042--3054}.
\bibitem[{Shuman et~al.(2013)Shuman, Narang, Frossard, Ortega and
  Vandergheynst}]{Shuman2013}
\bibinfo{author}{Shuman, D.I.}, \bibinfo{author}{Narang, S.K.},
  \bibinfo{author}{Frossard, P.}, \bibinfo{author}{Ortega, A.},
  \bibinfo{author}{Vandergheynst, P.}, \bibinfo{year}{2013}.
\newblock \bibinfo{title}{The emerging field of signal processing on graphs:
  Extending high-dimensional data analysis to networks and other irregular
  domains}.
\newblock \bibinfo{journal}{IEEE Signal Processing Magazine}
  \bibinfo{volume}{30}, \bibinfo{pages}{83--98}.
\bibitem[{Von~Luxburg(2007)}]{VonLuxburg2007AClustering}
\bibinfo{author}{Von~Luxburg, U.}, \bibinfo{year}{2007}.
\newblock \bibinfo{title}{A tutorial on spectral clustering}.
\newblock \bibinfo{journal}{Statistics and Computing} \bibinfo{volume}{17},
  \bibinfo{pages}{395--416}.
\bibitem[{Zong et~al.(2018)Zong, Song, Min, Cheng, Lumezanu, Cho and
  Chen}]{zong2018deep}
\bibinfo{author}{Zong, B.}, \bibinfo{author}{Song, Q.}, \bibinfo{author}{Min,
  M.R.}, \bibinfo{author}{Cheng, W.}, \bibinfo{author}{Lumezanu, C.},
  \bibinfo{author}{Cho, D.}, \bibinfo{author}{Chen, H.}, \bibinfo{year}{2018}.
\newblock \bibinfo{title}{Deep autoencoding gaussian mixture model for
  unsupervised anomaly detection}, in: \bibinfo{booktitle}{International
  Conference on Learning Representations}, pp. \bibinfo{pages}{1--19}.

\end{thebibliography}
